\documentclass[a4paper,fleqn]{cas-dc}
\usepackage[numbers,sort&compress]{natbib}
\usepackage{enumitem}
\usepackage{graphicx}
\usepackage{amsmath, amssymb, amsfonts}
\usepackage{booktabs}
\usepackage{array}
\usepackage{tabularx}
\usepackage{multirow}
\usepackage{xcolor}
\usepackage{caption}
\usepackage{subcaption}

\makeatletter
\def\Hy@Warning#1{}
\makeatother
\hypersetup{hypertexnames=false}
\makeatletter
\ExplSyntaxOn
\cs_set:Npn \__first_footerline: { }
\ExplSyntaxOff
\makeatother

\begin{document}
\let\WriteBookmarks\relax
\shorttitle{Reconfigurable Four-State Plasmonic Sensor}
\shortauthors{Ifty et al.}

\title[mode=title]{A Reconfigurable Four-State Sensor\\Based on Orthogonal Thermal\\and Chemical Stimuli}

\author[1]{Ashfaq Ur Rahman Ifty}[orcid=0009-0008-7292-0128]
\author[1]{Sadia Basher}[orcid=0009-0006-2684-1547]
\fnmark[1]
\author[1]{Abdullah Al Zayed}[orcid=0009-0003-4233-6899]
\fnmark[1]
\author[1]{Mohammad Abrar Kabir}[orcid=0009-0001-4187-2134]
\cormark[1]
\ead{abrarkabir@iut-dhaka.edu}
\author[1]{Rakibul Hasan Sagor}[orcid=0000-0002-8511-4466]

\affiliation[1]{organization={Department of Electrical and Electronic Engineering, Islamic University of Technology (IUT)},
                addressline={Gazipur-1704},
                country={Bangladesh}}

\cortext[cor1]{Corresponding author}
\fntext[fn1]{These authors contributed equally to this work.}

\begin{abstract}
This article presents a novel four-state refractive index (RI) sensor based on distinct phases of palladium (Pd) and vanadium dioxide (VO\textsubscript{2}), controlled by orthogonal thermal and chemical stimuli. This is a numerical, simulation-only study: all results are obtained from two-dimensional finite-element method (FEM) simulations. To the best of our knowledge, this is the first reported multi-state reconfigurable RI sensor tuned by external control. VO\textsubscript{2} serves as a phase change material (PCM) offering two thermally switchable phases with highly contrasting optical properties, while the hydrogen absorption property of Pd provides two additional phases. Combining these, four operational states are realized: (i)~$\alpha$-Pd/low-$T$ VO\textsubscript{2}; (ii)~$\beta$-PdH\textsubscript{x}/low-$T$ VO\textsubscript{2}; (iii)~$\alpha$-Pd/high-$T$ VO\textsubscript{2}; and (iv)~$\beta$-PdH\textsubscript{x}/high-$T$ VO\textsubscript{2}. Thermal crosstalk between Pd and VO\textsubscript{2} is mitigated through tungsten (W) doping of VO\textsubscript{2}. This work further introduces conditional state sensitivity (CSS), a novel figure of merit quantifying the optical contrast between each state transition. The proposed sensor achieves a peak sensitivity of 2871~nm/RIU for the undoped reference configuration and 2798~nm/RIU for the practical W-doped device, together with a peak CSS of 16{,}872~nm/RIU, establishing it as a multi-state reconfigurable platform with reduced thermal crosstalk and potential applications in industrial safety monitoring and optical logic gates.
\end{abstract}

\sloppy
\hfuzz=1000pt
{\hfuzz=\maxdimen\hbadness=10000\maketitle}

\section{Introduction}
\label{sec:intro}
\begin{figure*}
	\centering
	\includegraphics[width=.8\textwidth,height=0.82\textheight,keepaspectratio]{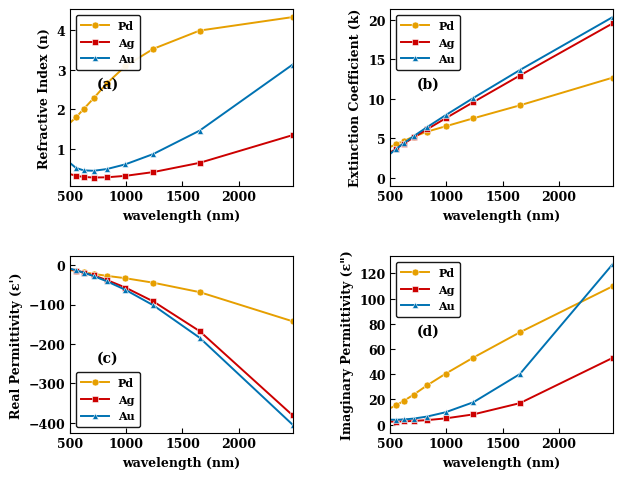}
	\caption{Wavelength dependent (a) refractive index, (b) extinction coefficient (c) real permittivity, and (d) imaginary permittivity of Pd, Ag, and Au \cite{werner2009optical}.}
	\label{pd}
\end{figure*}

Surface plasmon polaritons (SPPs) are quasiparticles 
originating from the resonant coupling between incident 
photons and the collective free-electron oscillations 
sustained at a metal--dielectric boundary~\cite{zheng2022thermal,meng2020selective}. Fundamentally, these polaritons represent
a coupled mode involving photons and surface charge density waves
\cite{law2004nanoribbon,krevsic2019electromagnetically,kabir2025high}. A defining characteristic
of SPPs is the strong confinement of the electromagnetic field at the boundary, which decays
exponentially into the adjacent media. This unique field localization allows SPPs to overcome
the conventional diffraction limit, enabling the precise manipulation and propagation of light
at the sub-wavelength scale \cite{li2018high,khani2018double}. In recent years,
SPP-based nanoscale sensors have gained considerable attention due to their ability to overcome
size limitations, leading to the creation of subwavelength devices such as Fabry--Perot
resonators \cite{wang2020towards}, Mach--Zehnder interferometers (MZIs), and refractive index
(RI) sensors \cite{al2021plasmonic}. A well-explored frontier is optical systems capable of
responding to multiple external stimuli simultaneously.

To achieve this, different routes were explored, such as creating active optical elements
\cite{zheludev2012metamaterials,duan2019magnesium,zhang2012photoinduced,wang2020towards}.
Common active tuning mechanisms include thermal tuning of vanadium dioxide
(VO\textsubscript{2}) \cite{boyce2022actively}, graphene array-driven tuning
\cite{mei2023operando}, micro-electromechanical systems (MEMS) mechanical tuning
\cite{oshita2020reconfigurable}, and tuning using palladium (Pd) \cite{sugawa2015refractive}.
However, nearly all reported designs remain fundamentally single-stimulus systems, offering
tunability along only one control axis at a time. Binary switching between two resonant
states, while useful, is insufficient for applications demanding nuanced, multi-level
classification of environmental conditions. While recent work has addressed crosstalk 
mitigation in dual-parameter optomechanical sensors through spatial separation of sensing 
regions \cite{taharat2025oxidation}, thermal crosstalk between chemically and thermally 
active materials in plasmonic systems remains largely unexplored. This motivates the search 
for sensor architectures that support orthogonal, independent control over multiple physical 
parameters simultaneously.

In this research, Pd and VO\textsubscript{2} were both used to achieve multiple independent
dimensions of active tuning, a combination that has not been investigated together in any
previous study of RI sensors. Pd is uniquely suited as a dual-function material: it serves
simultaneously as a plasmonic medium and a chemically active transducer, since hydrogen
(H\textsubscript{2}) absorption into its lattice drives a reversible $\alpha$-to-$\beta$
phase transition accompanied by pronounced changes in complex refractive index and
permittivity \cite{sharpe2025simultaneous,suzana2021structure,palm2018dynamic}.
VO\textsubscript{2}, on the other hand, undergoes a thermally driven metal--insulator
transition (MIT) near 68$^{\circ}$C, providing a second, thermally controlled optical
switching axis with large permittivity contrast \cite{wen2024review,qazilbash2007mott}.
The orthogonal nature of these two stimuli, chemical pressure governing Pd and temperature
governing VO\textsubscript{2}, enables the construction of a deterministic four-state
sensor space without ambiguity between control channels.

A critical practical challenge, however, is thermal crosstalk: elevated temperatures
required to switch VO\textsubscript{2} simultaneously accelerate H\textsubscript{2}
desorption from the Pd layer, destabilizing the $\beta$-PdH\textsubscript{x} phase and
rendering certain state combinations physically inaccessible in undoped systems
\cite{michalak2012uptake,schwarzer2022adsorption}. To resolve this, tungsten (W) doping
of VO\textsubscript{2} is employed to reduce the phase transition temperature from
68$^{\circ}$C to approximately 43$^{\circ}$C, significantly suppressing desorption kinetics
and enabling stable coexistence of the high-temperature VO\textsubscript{2} phase with the
$\beta$-PdH\textsubscript{x} phase \cite{bleu2023towards,soltani2004optical,sun2022room}.
This thermal engineering is essential for realizing all four states as genuinely accessible
operating regimes rather than theoretical constructs.

Using active tuning as a core mechanism, the RI sensor can be applied to industrial
protection, dual-analyte biosensing, and optical logic gates. In this article, orthogonal
control of PdH\textsubscript{x} alloy and VO\textsubscript{2} is introduced using
H\textsubscript{2} absorption-desorption and thermal manipulation. This control mechanism
creates novel four-state sensors with a combination of Pd--low-$T$ (VO\textsubscript{2}),
Pd--high-$T$ (VO\textsubscript{2}), PdH\textsubscript{x}--low-$T$ (VO\textsubscript{2}),
and PdH\textsubscript{x}--high-$T$ (VO\textsubscript{2}). A distinguishing feature of the
proposed device is its bidirectional spectral response: thermal stimuli drive a blue shift
through W-doped VO\textsubscript{2} \cite{lee2018isostructural,wen2024review}, while
H\textsubscript{2} absorption drives an opposing red shift through lattice expansion of
PdH\textsubscript{x} \cite{suzana2021structure,sharpe2025simultaneous}. This spectral
orthogonality allows unambiguous classification of environmental states into safe,
thermal warning, gas leak, and critical danger, a capability unattainable
with single-stimulus binary sensors.

For the structural implementation, the metal--insulator--metal (MIM) waveguide configuration
is preferred due to its compact node size \cite{zhang2013fano}, superior electromagnetic
(EM) confinement \cite{chouChao2023}, and a favorable balance between propagation length
and loss \cite{zhang2021formation}. In this research, Pd and VO\textsubscript{2} were
implemented in a MIM configuration, enabling better light confinement and resonance. A ring
cavity coupled with a waveguide and a central nanostrip was designed, achieving a remarkable
maximum conditional state sensitivity (CSS) of 16{,}872~nm/RIU and a peak sensitivity of
2{,}871~nm/RIU for the undoped reference configuration (2{,}798~nm/RIU for the W-doped
device). CSS is introduced in this work as a novel performance metric that quantifies
the effective spectral contrast between each pair of states, providing a more complete
picture of multi-state sensor discrimination than conventional single-state sensitivity
alone. Qu et al.~\cite{qu2024mim} proposed a dual-stub MIM waveguide 
structure exploiting independently tunable Fano resonances, 
achieving peak sensitivities of 2780 and 3580~nm/RIU across 
two channels. Chen et al.~\cite{chen2024srrr} demonstrated a 
semicircular ring rectangular resonator (SRRR) coupled to a 
baffled MIM waveguide, yielding a maximum sensitivity of 
2560~nm/RIU with a figure of merit of 1080~RIU$^{-1}$. 
Yang et al.~\cite{yang2024nano} presented a triple-disk 
T-shaped cavity (STDTC) MIM nanosensor achieving a sensitivity 
of 3020~nm/RIU and a quality factor of 51.89. Chou Chao 
et al.~\cite{chouChao2023} proposed a multichannel 
elliptical-ring MIM sensor supporting five Fano resonance modes 
with a maximum sensitivity of 4500~nm/RIU for refractive index 
sensing. The proposed sensor competes with these benchmarks while
additionally offering four-state logic capability and dual-stimulus reconfigurability
that none of the prior works provide.

In industrial environments, safety challenges such as explosions due to highly flammable
H\textsubscript{2} and overheating in battery packs have driven significant interest in
sensors leveraging surface plasmon resonance (SPR) and MIM waveguides, particularly due
to their immunity to electromagnetic interference (EMI) \cite{luong2021sub,mei2023operando}.
These tasks traditionally required separate sensors, but using multiple states of this
reconfigurable sensor, they can be implemented using one compact device. This study
demonstrates that nanoscale MIM sensors can be engineered for multi-dimensional active
control, opening new directions in reconfigurable plasmonic sensing, optical logic, and
compact industrial safety monitoring that were not explored in previous research.
\section{Material Selection}
\label{sec:theory}

\subsection{Rationale for Pd and Pd-Hydrogen Alloy}

Palladium (Pd) is widely known as one of the most suitable metals for H\textsubscript{2} 
absorption or desorption; hence, it is used for hydrogen sensing, as larger amounts of 
H\textsubscript{2} can be absorbed into its lattice, forming a bulk hydride 
\cite{sterl2020design}. Hydrogen dissociates into atoms that absorb on the Pd surface; 
these adsorbed atoms then diffuse into near-surface subsurface sites and ultimately 
penetrate deeper to occupy interstitial positions within the bulk lattice 
\cite{suzana2021structure}. The hydride formation in palladium is largely reversible: upon 
reducing the ambient hydrogen pressure or increasing temperature, dissolved hydrogen 
diffuses back to the surface and desorbs, allowing the Pd--H alloy to decompose and the 
lattice to relax toward its original state \cite{li2009preparation}. The palladium hydride 
alloy (PdH\textsubscript{x}) exhibits two crystalline 
phases named $\alpha$ and $\beta$ (also known as $\alpha'$) at room temperature. It is 
identified as pure $\alpha$-phase when $x<0.017$; with more covalent Pd--H bonding, pure 
$\beta$ is formed with more metallic character when $x>0.58$ \cite{sharpe2025simultaneous}. 
The Pd crystalline structure expands by approximately 3.5\% while 
transforming from the $\alpha$ phase to the $\beta$ phase, reflecting the incorporation of a high 
hydrogen concentration and the formation of a crystalline structure \cite{suzana2021structure}. 
In practical hydrogen-sensing operation, this transition is exploited as a reversible change 
between hydrogen-depleted and hydrogen-loaded PdH\textsubscript{x} states within $0<x<1$, 
rather than as a fixed stoichiometric compound such as PdH\textsubscript{2}.
The associated changes in lattice constant and electronic structure lead to pronounced 
variations in the complex refractive index $n(\lambda)$ and extinction coefficient 
$k(\lambda)$, and hence in the complex permittivity $\varepsilon(\lambda)$. These 
composition-dependent optical properties provide the physical basis for active tuning of 
surface-plasmon-polariton (SPP) resonances in Pd-based hydrogen sensors.
When an optical excitation is given, the photons interact with the electrons 
of the metal and result in a collective oscillation \cite{zhang2012surface}. This surface wave can be used to manipulate the propagation 
and dispersion of light at the nanometer scale \cite{law2004nanoribbon, 
khani2018double,taharat2025advancing}. When exposed to light, the complex permittivity 
($\hat{\varepsilon}$) of Pd changes with the frequency of the incident 
radiation. This phenomenon can be modeled by the Lorentz--Drude framework 
\cite{rakic1998optical}, which is expressed as follows:
\begin{equation}
    \hat{\varepsilon}_{\text{Pd}}(\omega) = 1 - \frac{\Omega_p^2}{\omega^2 - i\omega\Gamma} + \sum_{x=1}^{N} \frac{f_x \omega_p^2}{(\omega_x^2 - \omega^2) + i\omega\Gamma_x},
    \label{eq:lorentz_drude}
\end{equation}
\begin{figure}
\centering
\includegraphics[width=0.8\linewidth,height=0.44\textheight,keepaspectratio]{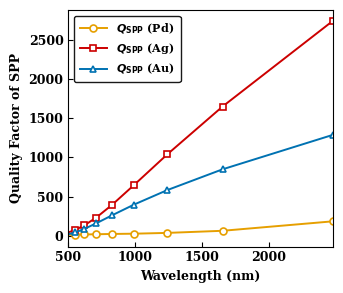}
    \caption{\textit{Qspp curve of Ag, Au and Pd.}}
    \label{Qspp}
\end{figure}
\begin{figure}
    \centering
    \includegraphics[width=\linewidth,height=0.44\textheight,keepaspectratio]{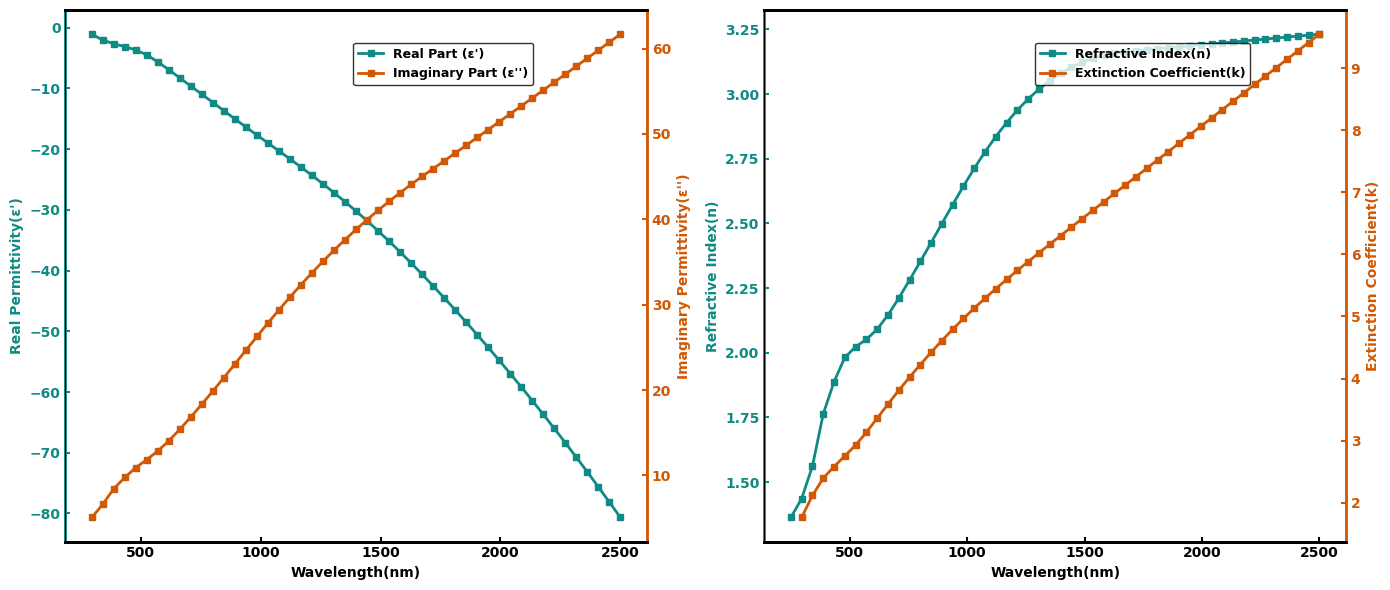}
    \caption{(a) Real and imaginary permittivity of Pd--H alloy and (b) refractive index and extinction coefficient of Pd--H alloy \cite{palm2018dynamic}.}
    \label{pdh}
\end{figure}

The complex refractive index $(n, k)$ values for palladium (Pd), silver (Ag), and gold (Au) 
were adopted from the experimental data of Werner et al.\ \cite{werner2009optical} over 
the 0.248--12.4~\textmu m spectral range. To evaluate the plasmonic performance of these 
materials, we derived the complex permittivity and the Surface Plasmon Polariton (SPP) 
quality factor of SPP ($Q_\text{SPP}$) using the following relations:
\begin{gather}
    \varepsilon = \varepsilon' + i\varepsilon'' = (n + ik)^2 \tag{2} \\
    Q_\text{SPP} = \frac{(\varepsilon')^2}{\varepsilon''} \tag{3}
\end{gather}

Figure~\ref{pd} illustrates the comparative permittivity and Fig.~\ref{Qspp} illustrates the $Q_{\text{SPP}}$ 
for the three metals. As expected, Pd exhibits a significantly lower $Q_\text{SPP}$ 
compared to the noble metals (Ag and Au) in the near-infrared, indicating higher Ohmic 
losses. However, this plasmonic penalty is offset by the functional utility of Pd. While 
the dielectric functions of Ag and Au are essentially static and chemically inert, Pd 
allows for dynamic optical modulation. Hydrogen absorption induces changes in both $n$ and 
$k$ (and consequently in the real and imaginary permittivity), enabling the active tuning of the SPP resonance.
Therefore, in this design, Pd serves dual roles as both the plasmonic medium and 
the active sensing transducer, whereas Ag and Au are presented solely as high-quality 
plasmonic references. Figure~\ref{pdh} shows the real and 
imaginary permittivity, refractive index, and extinction coefficient of the Pd--H alloy under 
7~bar H\textsubscript{2} gas pressure. The experimental data were taken from Palm et al.\ 
\cite{palm2018dynamic}.

\subsection{Rationale for Vanadium Dioxide (VO\textsubscript{2}) as Phase Change Material}

\begin{figure*}
	\centering
	\includegraphics[width=.75\textwidth,height=0.82\textheight,keepaspectratio]{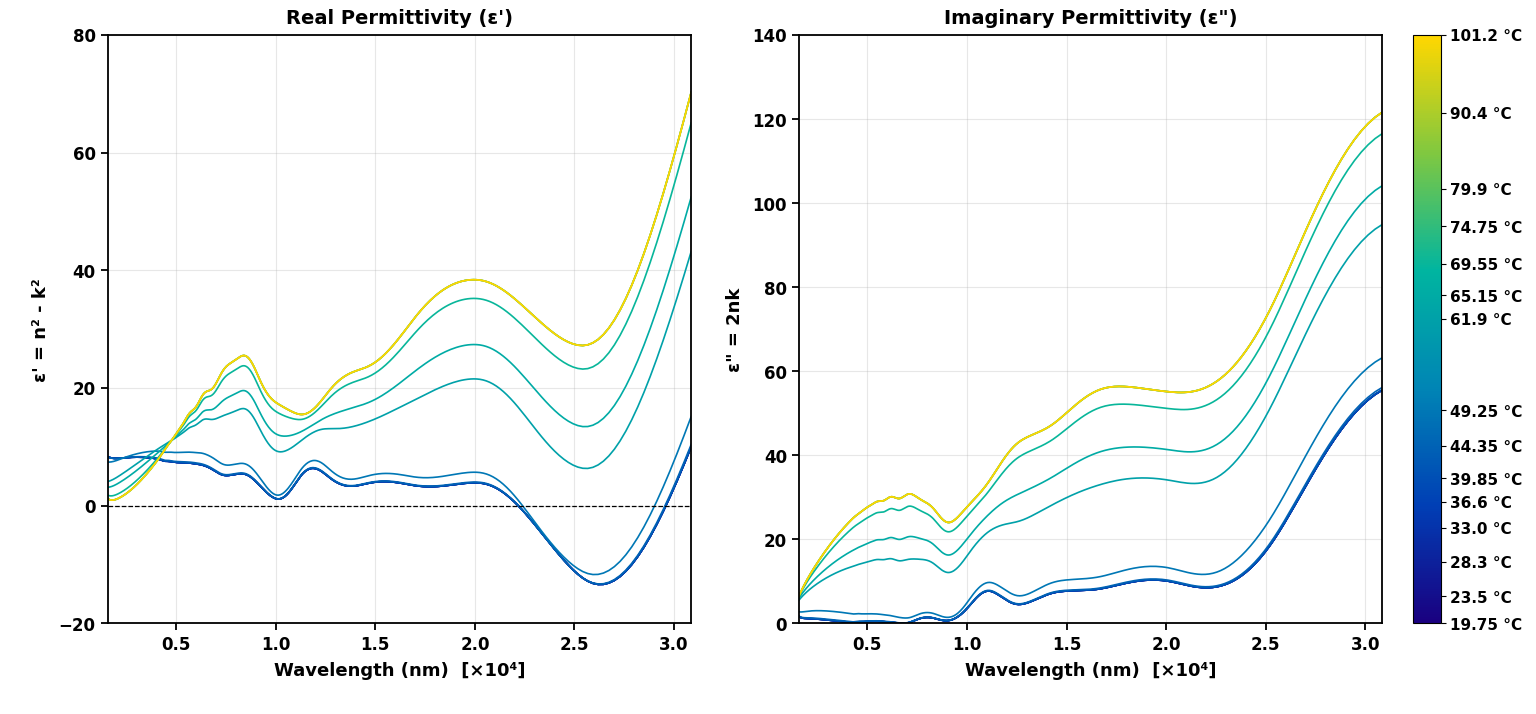}
\caption{\textit{Temperature-dependent permittivity of undoped VO\textsubscript{2} redrawn from Sun et al.\ \cite{sun2022room}. (a) Real permittivity and (b) imaginary permittivity as a function of wavelength for temperatures between 20$^\circ$C and 100$^\circ$C.}}
    \label{fig:vo2_permittivity}
\end{figure*}

Vanadium dioxide (VO\textsubscript{2}) exhibits a thermally driven metal--insulator 
transition (MIT) at around 68$^\circ$C, accompanied by a large change in its complex 
permittivity \cite{wen2024review}. This combination of near-ambient switching temperature, 
substantial optical contrast, and compatibility with thin-film growth makes VO\textsubscript{2} 
an attractive phase-change material for actively tunable photonic and plasmonic devices 
\cite{wen2024review}. Figure~\ref{fig:vo2_permittivity} shows that the real permittivity changes sign as the temperature is increased from 49.25$^\circ$C to 69.55$^\circ$C.

At low temperatures, VO\textsubscript{2} is in the monoclinic insulating M1 phase, where 
V--V dimerization opens a bandgap of roughly 0.6~eV and the material behaves as a 
semiconductor with positive real permittivity in the near-/mid-infrared. When the 
temperature exceeds the transition point ($\approx$68$^\circ$C for undoped VO\textsubscript{2}), 
the lattice transforms to a tetragonal rutile R phase, the V--V dimers break, the bandgap 
closes, and VO\textsubscript{2} becomes metallic with negative real permittivity and 
increased optical loss over a broad spectral range \cite{qazilbash2007mott}. Volatile PCMs 
like VO\textsubscript{2} revert to their original phase naturally upon stimulus withdrawal, 
enabling fast, continuous optical modulation critical for dynamic plasmonic devices 
\cite{oguntoye2023continuously}. The structural phase transition (SPT), changing from a 
monoclinic structure at low temperatures to a tetragonal rutile lattice at high 
temperatures, is governed by strong electron--electron correlations (Mott--Hubbard 
mechanism) and structural distortions (Peierls instability) \cite{oguntoye2023continuously}.

\subsection{Rationale for W-Doping of VO\textsubscript{2}}

\begin{figure*}
    \centering
    \includegraphics[width=.75\textwidth,height=0.82\textheight,keepaspectratio]{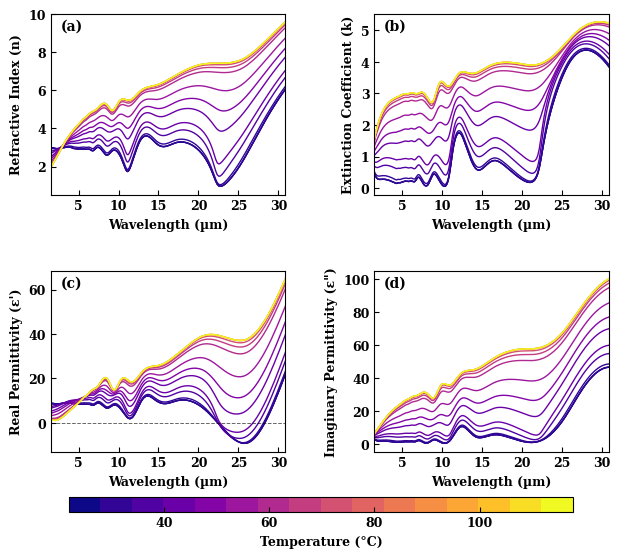}
    \caption{Temperature-dependent complex refractive index of W-doped VO\textsubscript{2} in the mid-infrared (1--32~\textmu m), color-coded by temperature. (a) $n(\lambda)$, (b) $k(\lambda)$ redrawn from Sun et al.\ \cite{sun2022room}; (c) real permittivity and (d) imaginary permittivity derived from Eq.~(2).}
    \label{wvo2}
\end{figure*}

The phase transition temperature of undoped VO\textsubscript{2} thin film is around 
68$^\circ$C, making it nonideal for our sensor, which is targeted at lower 
temperatures \cite{bleu2023towards}. Substitutional incorporation of W$^{6+}$ at V$^{4+}$ 
lattice sites has been extensively studied as a route 
to suppressing the phase transition temperature of 
VO$_2$ thin films, with concurrent modifications 
to their thermochromic response and surface 
morphology. Sun et al.\ \cite{sun2022room} showed that W doping of VO\textsubscript{2} lowers the phase transition temperature towards room 
temperature. Soltani et al.\ \cite{soltani2004optical} prepared 1.6\% W-doped 
VO\textsubscript{2} thin films using a W-doped vanadium oxide sputtering target, achieving 
a transition temperature of 36$^\circ$C and demonstrating all-optical switching at the 
telecom wavelength of 1.55~\textmu m. Figure~\ref{wvo2} illustrates the 
temperature-dependent optical constants of the 0.93\% W-doped film.

For this work, a 
doping concentration of 0.93\% was chosen for optimal operation of all four states, considering the hydrogen desorption of the PdH alloy. The experimental dataset of 0.93\% doped VO\textsubscript{2} was 
taken from Sun et al.\ \cite{sun2022room}. Our chosen operating temperature is 
43$^\circ$C. As the temperature increases from 25$^\circ$C to 100$^\circ$C, the material 
undergoes a gradual insulator-to-metal transition (IMT). Unlike the abrupt binary switching 
seen in some bulk PCMs, the W-doped film exhibits a continuous evolution of its refractive 
index ($n$) and extinction coefficient ($k$) (Figures~\ref{wvo2}(a) 
and~(b)). This graded response allows for fine analog tuning of the plasmonic resonance.

This strong thermo-optic modulation serves as the primary mechanism 
for the active control of the proposed sensor's spectral response.

The proposed sensor operates within the 2.0--2.5~\textmu m spectral window. As illustrated 
in Fig.~\ref{wvo2}, the film retains positive real permittivity across the entire 
temperature range within this band. This indicates 
that, unlike in the far-infrared, where VO\textsubscript{2} behaves as a Drude metal, the 
material remains in the optical dielectric regime at near-infrared frequencies, even after 
the phase transition. However, the absence of a negative permittivity ``metallic'' response 
does not hinder device performance. Effective active switching does not require a full 
dielectric-to-metal transition; rather, it relies on the substantial optical contrast 
between the two thermal states:
\begin{enumerate}
    \item \textbf{Low-T VO\textsubscript{2} (Insulating-like):} Characterized by lower 
    refractive index and extinction.
    \item \textbf{High-T VO\textsubscript{2} (Lossy Dielectric):} Characterized by 
    significantly elevated $n$ and $k$.
\end{enumerate}
The transition between these two 
distinct dielectric states induces a massive shift in the optical path length and absorption 
of the plasmonic mode. Therefore, the device exploits the large differential refractive 
index and extinction contrast generated by the structural phase change, creating a binary 
switching functionality defined effectively by the low-T and high-T optical phases.

\section{Structural Design and Methodology}
\subsection{Geometrical Structure}

\begin{figure*}
\centering
    \includegraphics[width=0.7\textwidth,height=0.82\textheight,keepaspectratio]{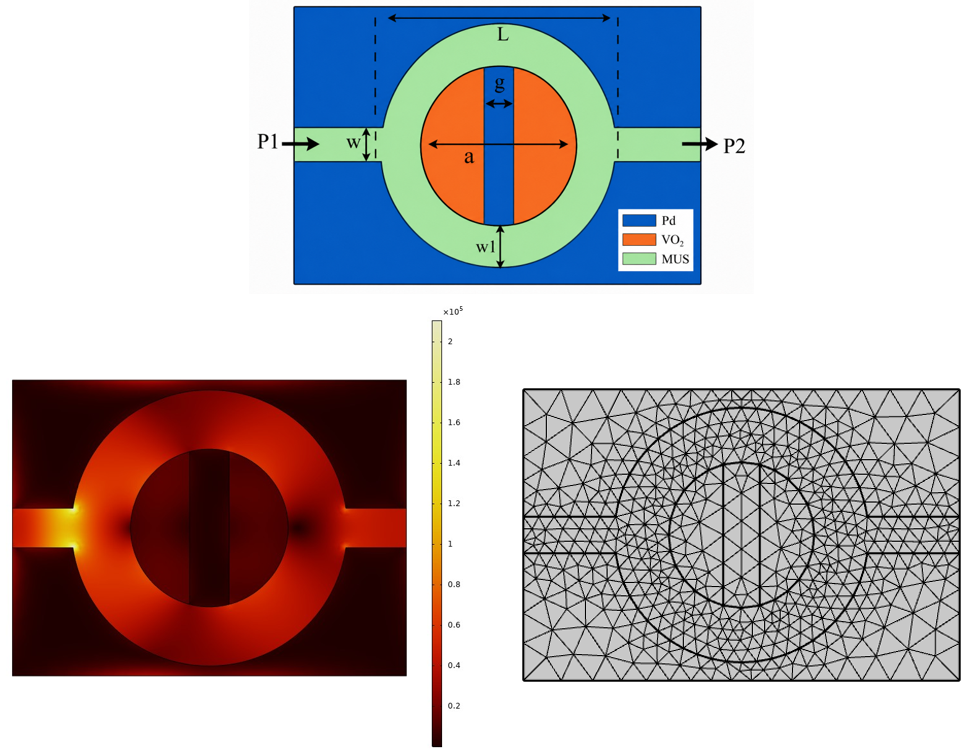}
    \caption{\textit{(a)~Schematic diagram of the proposed sensor. (b)~Electric field distribution. (c)~Meshing of the structure}}
    \label{structure}
\end{figure*}

Figure~\ref{structure}(a) shows the schematic diagram of the proposed sensor. Following the 
material nomenclature of the fabrication sequence in Fig.~\ref{fabrication}, the surrounding 
cladding and the MIM bus waveguide are formed in Pd, the ring resonator is patterned in 
W-doped VO\textsubscript{2}, and a central Pd nanostrip is defined inside the cavity. The 
remaining open volume enclosed by the ring is filled by the material under sensing (MUS). 
Table~\ref{tab:geometric_params} 
provides the geometric parameter values for the designed sensor. The W-doped 
VO\textsubscript{2} ring and the central Pd nanostrip are side-coupled to the bus waveguide. 
The introduction of this 
nanostrip introduces additional parameters for design tunability and optimization. The 
integration of VO\textsubscript{2} enables multistate operations together with the Pd layers. 
The electric field distribution at the resonant wavelength has been shown 
in Fig.~\ref{structure}(b).

TM-polarized light enters the input port (P1), and the total reflectance of the structure is 
monitored over the 1.5--3.5~\textmu m wavelength range; all spectra reported in this work 
are total reflectance spectra. The mechanism is stimulated by 
external catalysts such as hydrogen pressure and temperature in the surroundings. For this 
research, free triangular meshing has been used with finer resolution in the waveguide and 
inner cavities and coarser discretization in the metal surface, which is shown in 
Fig.~\ref{structure}(c). The outer domain boundaries are terminated with scattering 
boundary conditions, which absorb outgoing electromagnetic radiation and prevent spurious 
reflections back into the computational region. Input and output ports of the waveguide are 
defined with port boundary conditions to excite and monitor the fundamental TM-polarized 
SPP mode. This mesh density ensures convergence of the computed reflectance spectra 
across the 1500--3500~nm operating wavelength range.

\subsection{Theoretical Analysis}
\begin{table}
\caption{Initial Geometric Parameters of the Proposed Sensor}
\label{tab:geometric_params}
\centering
\begin{tabular}{lcc}
\hline
\textbf{Parameter} & \textbf{Notation} & \textbf{Initial Value (nm)} \\
\hline
Waveguide width      & $w$   & 100 \\
Ring diameter        & $L$   & 700 \\
Central nanostrip width & $g$ & 100 \\
Ring width           & $w_1$ & 150 \\
Inner cavity (MUS) diameter & $a$ & 400 \\
\hline
\end{tabular}
\end{table}
The optical response is computed by solving the frequency-domain Maxwell equations:
\begin{equation}
    \nabla \times (\mu_r^{-1}\nabla \times \mathbf{E}) - k_0^2\varepsilon_r\mathbf{E} = 0
    \tag{4}
\end{equation}
where
\begin{equation}
    k_0 = \frac{2\pi}{\lambda}
    \tag{5}
\end{equation}
is the free-space wavenumber, and $\varepsilon_r$ and $\mu_r$ are the relative permittivity 
and permeability tensors. Simulations are performed using the finite element method (FEM) 
in COMSOL Multiphysics Electric Fields and Waves module in 2D geometry with TM 
polarization. A 2D model provides considerable time and resource savings compared to 3D 
simulations. The computational cost of full 3D simulation is avoided 
without sacrificing accuracy: prior experimental work 
confirms that the 2D FEM model remains predictive for 
structures where the out-of-plane dimension exceeds 
800~nm~\cite{al2021highly}.  Given that the waveguide 
width is sub-wavelength relative to the excitation, only 
the fundamental TM mode is supported. The corresponding 
SPP dispersion relation governing this mode in the MIM 
configuration takes the form
\begin{equation}
    \varepsilon_d k_m + \varepsilon_m k_d \tanh\left(\frac{w k_d}{2}\right) = 0
    \tag{6}
\end{equation}
where $\varepsilon_d$ and $\varepsilon_m$ are the relative permittivity of the dielectric 
and metal, respectively. $k_m$ and $k_d$ are defined by momentum conservation as follows:
\begin{equation}
    k_{m,d}^2 = \beta^2 - \varepsilon_{m,d}k^2
    \tag{7}
\end{equation}
and
\begin{equation}
    k = \frac{2\pi}{\lambda}
    \tag{8}
\end{equation}
where $\beta$ denotes the SPP propagation constant and $k$ 
the free-space wave vector. Applying these dispersion 
relations, the real part of $n_\text{eff}$ and the SPP 
propagation length $L_\text{SPP}$ can be extracted for 
each of the four operational states. The resonant wavelength 
$\lambda_\text{res}$ of the cavity follows directly from 
standing-wave quantization, whereby $\lambda_\text{res}$ 
scales with $\text{Re}(n_\text{eff})$ according 
to~\cite{chou2021significantly}:

\begin{equation}
    \lambda_\text{res} = \frac{\text{Re}(n_\text{eff}) \cdot L_\text{SPP}}
    {m - \dfrac{\phi_\text{ref}}{\pi}}
    \tag{9}
\end{equation}

\subsection{Operating Principle and State Definition}

The configurable four-state sensor operates independently by modulating two orthogonal 
stimuli, as discussed previously. Table~\ref{tab:control_parameters} summarizes all control 
parameters for each state.

\begin{table*}
    \centering
    \caption{\textbf{Control parameters for each state.}}
    \small
    \setlength{\tabcolsep}{4pt}
    \begin{tabularx}{\textwidth}{@{} l >{\raggedright\arraybackslash}X 
    >{\centering\arraybackslash}p{1.6cm} c >{\raggedright\arraybackslash}p{2.2cm}
    >{\raggedright\arraybackslash}p{2.2cm} >{\raggedright\arraybackslash}p{1.3cm} @{}}
        \toprule
        \textbf{State} & 
        \textbf{Temperature Regime} & 
        \textbf{H$_2$ Press. (atm)} & 
        \textbf{Pd Phase} & 
        \textbf{VO$_2$ Optical Phase} & 
        \textbf{Control Stimulus} & 
        \textbf{H/Pd Ratio} \\
        \midrule
        \textbf{I or 00}   
            & Lower than Critical Temp. \cite{michalak2012uptake}  
            & $<0.01$ \cite{sharpe2025simultaneous} 
            & $\alpha$-Pd \cite{michalak2012uptake} 
            & Low-$T$ (Insulating-like) \cite{lee2018isostructural} 
            & Reference             
            & 0.017 \cite{sharpe2025simultaneous} \\
        \textbf{II or 01}  
            & Lower than Critical Temp. \cite{michalak2012uptake}  
            & 2--5 \cite{sharpe2025simultaneous}    
            & $\beta$-PdH \cite{michalak2012uptake} 
            & Low-$T$ (Insulating-like) \cite{lee2018isostructural} 
            & Chemical (Pressure)   
            & 0.58 \cite{sharpe2025simultaneous}  \\
        \textbf{III or 10} 
            & Higher than Critical Temp. \cite{michalak2012uptake} 
            & $<0.01$ \cite{sharpe2025simultaneous} 
            & $\alpha$-Pd \cite{michalak2012uptake} 
            & High-$T$ (Lossy dielectric) \cite{lee2018isostructural}       
            & Thermal (Temperature) 
            & 0.017 \cite{sharpe2025simultaneous} \\
        \textbf{IV or 11}  
            & Higher than Critical Temp. \cite{michalak2012uptake} 
            & 2--5 \cite{sharpe2025simultaneous}    
            & $\beta$-PdH \cite{michalak2012uptake} 
            & High-$T$ (Lossy dielectric) \cite{lee2018isostructural}       
            & Combined (T + P)      
            & 0.58 \cite{sharpe2025simultaneous}  \\
        \bottomrule
    \end{tabularx}
    \label{tab:control_parameters}
\end{table*}

Thermal stimuli control the phase transition of vanadium dioxide. Chemical stimuli 
(mostly comprising hydrogen pressure and H/Pd ratio) control the Pd--H phase. In the 
$\alpha$-phase, Pd--H has low hydrogen content ($x < 0.017$), and in the $\beta$-phase, 
Pd--H contains high hydrogen content ($x > 0.58$). The intermediate regime 
($0.017 < x < 0.58$) exhibits $\alpha$--$\beta$ coexistence with pronounced hysteresis 
during loading--unloading cycles.

Surface plasmon polaritons (SPPs) in the MIM waveguide arise from the coupling of 
TM-polarized light (magnetic field perpendicular to the plane) at the metal--dielectric 
interfaces. The fundamental operating mechanism relies on how the optical permittivity 
$\varepsilon(\omega)$ of each layer changes across the four states. Changes in the complex 
permittivity of Pd--H and VO\textsubscript{2} directly modulate (i) the effective mode 
index, (ii) the spatial confinement of electromagnetic fields in the sensing region, and 
(iii) the propagation loss. This allows each state to exhibit a unique reflectance spectrum, 
making the device inherently tunable and configurable without geometric modification.

The complete transition to the rutile phase is not 
thermodynamically accessible within the proposed operating limits. However, the spectral 
curves in Section~\ref{subsec:results} indicate that the temperature-dependent permittivity change is 
significant. This functional phase variation is adequate to drive the multi-state logic, 
confirming the utility of VO\textsubscript{2} as the switching element in the proposed 
design. Tungsten doping of VO\textsubscript{2} is essential for solving crosstalk but it does not help VO\textsubscript{2} to operate in distinct 
rutile and monoclinic phases. Consequently, to distinguish the operating regimes for 
tungsten-doped VO\textsubscript{2}, the VO\textsubscript{2} film at the baseline 
temperature is herein designated as VO\textsubscript{2} (Low-T), while the film at the 
elevated temperature is referred to as VO\textsubscript{2} (High-T). Therefore, the device 
operates across four distinct regimes: State~I ($\alpha$-Pd / Low-T VO\textsubscript{2}), 
State~II ($\beta$-PdH / Low-T VO\textsubscript{2}), State~III ($\alpha$-Pd / High-T 
VO\textsubscript{2}), and State~IV ($\beta$-PdH / High-T VO\textsubscript{2}).

Electric field distributions were simulated at resonance for all four material states to 
validate the sensor's operational principle. All states exhibit strong electromagnetic 
field confinement within the cavity together with a pronounced reflectance minimum, 
confirming that these minima correspond to genuine resonant modes, which can be seen in 
Fig.~\ref{structure}(b). The spatial field distribution shows pronounced localization in the 
Pd/PdH and VO\textsubscript{2} regions, creating hotspots where field intensity exceeds 
10 times the background level, the physical basis for high sensitivity. The fundamental 
resonant mode profile remains stable across all four states ($>$95\% modal overlap), 
demonstrating that spectral tuning arises from material property changes rather than 
excitation of different modes. The wavelength shift $\Delta\lambda$ between states 
constitutes the primary sensing observable. This analysis validates that the sensor 
architecture achieves strong field confinement, efficient coupling to tunable materials, 
and stable multi-state operation for reliable sensor discrimination.

\subsection{Universal State Transition Limitations}

While Table~\ref{tab:control_parameters} illustrates the theoretical framework for 
orthogonal control, not all state-to-state transitions are directly achievable due to the 
inherent material properties of the Palladium--Hydrogen (Pd--H) alloy and undoped Vanadium 
Dioxide (VO\textsubscript{2}). Although the proposed universal control mechanism 
theoretically allows for immediate switching between any two states, material limitations, 
i.e., thermal crosstalk, make certain direct transitions impractical in practice.

Specifically, transitions between States~I and~III, as well as States~II and~IV, are 
thermally driven and reversible. Conversely, transitions between States~I and~II and 
States~III and~IV are governed by hydrogen pressure modulation. However, in an undoped 
system the diagonal transition from State~I to State~IV is impractical; it requires 
simultaneous increases in both pressure and temperature, but elevated temperatures trigger 
hydrogen desorption in the Pd layer, hindering the system from transitioning to State~IV 
\cite{sharpe2025simultaneous}. As shown in Section~\ref{subsec:crosstalk}, tungsten doping 
lowers the VO\textsubscript{2} transition temperature so that the simultaneous application 
of temperature and pressure required for State~IV becomes physically sustainable. This 
stabilization is particularly significant for 
the transition between State~III ($\alpha$-Pd / High-T VO\textsubscript{2}) and State~IV 
($\beta$-PdH / High-T VO\textsubscript{2}). This transition is categorized as highly 
complex because it represents a ``double-stimuli'' shift; it requires the system to maintain 
its elevated thermal state (keeping VO\textsubscript{2} in the high-T regime) while 
successfully navigating the $\alpha$-to-$\beta$ phase transformation in the Pd layer. In 
undoped systems, the heat required for State~III would catalyse immediate hydrogen 
rejection, making it impossible to ``cross over'' into State~IV.

\subsection{Crosstalks and Solutions}
\label{subsec:crosstalk}

Thermal crosstalk between Pd--H layers and VO\textsubscript{2} exhibits a significant 
challenge, as heating VO\textsubscript{2} in the phase transition region directly affects 
hydrogen absorption/desorption, as Michalak et al.\ \cite{michalak2012uptake} demonstrated 
with temperature-programmed desorption (TPD), showing the H\textsubscript{2} desorption 
rate increases with $T$. Hydrogen desorption from the Pd layer fundamentally limits 
proposed device functionality and reversibility. The temperature-dependent desorption rate 
constant is described by the concentration-dependent Arrhenius Eq.~(10) 
\cite{michalak2012uptake}. Also, the activation energy for surface desorption follows a 
nonlinear coverage dependence shown in Eq.~(11) \cite{michalak2012uptake}.
\begin{equation}
    k_\text{des}(T, C_\text{SS}) = \nu_0 \times \exp\!\left(
    -\frac{\Delta E_\text{des}(C_\text{SS})}{k_B T}\right) \tag{10}
\end{equation}
\begin{equation}
    \Delta E_\text{des}(C_\text{SS}) = 0.42 + 0.25\theta + 0.74\left(1 - 
    \exp\!\left[-\frac{C_\text{SS}^2}{0.3}\right]\right)~\text{eV} \tag{11}
\end{equation}
Here $\nu_0 = 10^{13}~\text{s}^{-1}$ is the pre-exponential factor, and $k_B$ is the 
Boltzmann constant. $\theta$ is surface coverage, and $C_\text{SS}$ is the subsurface 
hydrogen concentration. This kinetic model reveals that subsurface hydrogen dramatically 
lowers the desorption barrier by up to 0.74~eV \cite{schwarzer2022adsorption}, which 
enables rapid hydrogen loss even at moderate temperatures. At higher temperatures, the 
Pd--H alloy can be said to be nonexistent or ineffective. In an undoped system, therefore, 
the coexistence of rutile VO\textsubscript{2} and $\beta$-phase PdH\textsubscript{x} is not 
feasible, which would restrict the fourth state to a purely theoretical regime.

Tungsten doping provides an effective route to lower the VO\textsubscript{2} phase 
transition temperature, which is essential for stabilizing State~IV; however, this 
necessitates a nomenclature shift where the material is designated as VO\textsubscript{2} 
(high-T) rather than stoichiometric rutile VO\textsubscript{2}. Temperature dependency 
with W doping can be shown using the following Eq.~(12) demonstrated by Bleu et 
al.\ \cite{bleu2023towards}:
\begin{equation}
    T_t = -25.70 \times \text{at.\% W} + 67.13 \tag{12}
\end{equation}
To resolve the thermodynamic conflict between the required phase transition and hydrogen 
stability, tungsten (W) doping was employed. The substitution of V\textsuperscript{4+} 
with W\textsuperscript{6+} ions injects excess charge carriers (electrons) into the 
lattice, destabilizing the low-temperature monoclinic phase and inducing a linear 
reduction in the critical temperature ($T_c$) \cite{tan2012unraveling,takami2012filling}. 
By precisely tuning the doping concentration, the transition temperature was lowered to 
43$^\circ$C. The doping concentration was chosen to be 0.93\% W.

This thermal shift is critical for enabling the previously infeasible state transitions. 
In the undoped VO\textsubscript{2} system ($T_c = 68^\circ$C), the transition to State~IV 
(high temperature, high pressure) fails because the elevated temperature triggers rapid 
hydrogen desorption, effectively destroying the ``high pressure'' condition required for 
the $\beta$-PdH phase. However, tungsten doping of VO\textsubscript{2} reduces the phase 
transition temperature to approximately 43$^\circ$C. At this reduced $T_c$ of 43$^\circ$C, 
the lower operating temperature significantly suppresses hydrogen desorption kinetics 
compared to the 68$^\circ$C regime, enabling stable maintenance of the high-pressure 
$\beta$-PdH phase while simultaneously accessing the VO\textsubscript{2} (High-T) phase 
\cite{michalak2012uptake,schwarzer2022adsorption}. Critical temperature 
decrement by tungsten doping is governed by Eq.~(12), as demonstrated by Bleu et 
al.\ \cite{bleu2023towards}. This creates a stable thermodynamic window where the 
VO\textsubscript{2} can switch between the VO\textsubscript{2} (High-T) and 
VO\textsubscript{2} (Low-T) phases without forcing the desorption of hydrogen from the 
palladium layer. Consequently, the simultaneous application of temperature and pressure 
required for State~IV becomes physically sustainable, rendering all control paths, 
including the complex transitions between States~III and~IV, fully operational. By 
lowering the thermal threshold via W-doping, the ``distance'' between the chemical and 
thermal equilibrium points is reduced. This allows for a synchronized transition where 
hydrogen loading can occur without being countered by high-temperature desorption 
kinetics. As a result, all control paths, including this previously hindered diagonal 
transition, are rendered fully operational.
\begin{figure*}
    \centering
    \includegraphics[width=.75\textwidth,height=0.82\textheight,keepaspectratio]{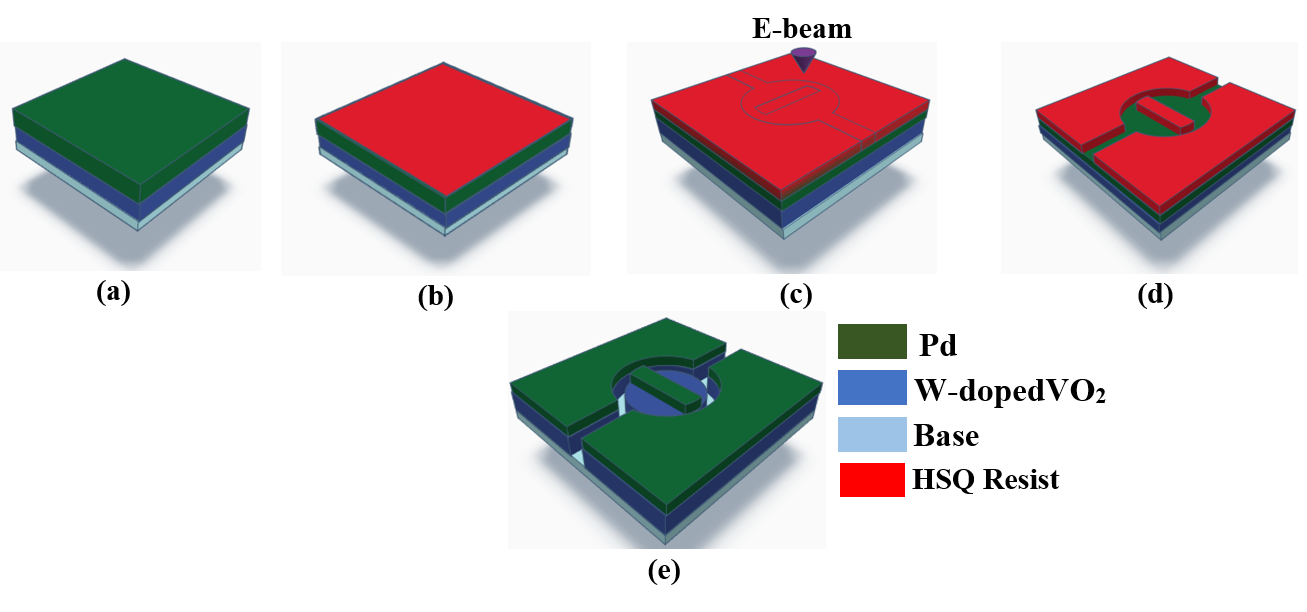}
    \caption{Fabrication process. (a)~SOI wafer, (b)~EBL resist coating, (c)~Electron beam exposure, (d)~Reactive ion etching, and (e)~Resist lift-off.}
    \label{fabrication}
\end{figure*}
Here $T_t$ is the transition temperature and \%W is the percentage of W doping. 
Figure~1(b) of Bleu et al.\ \cite{bleu2023towards} illustrated an inverse relation of 
transition temperature and tungsten doping percentage. The optimal operating temperature 
for hydrogen storage in palladium is determined by the phase stability of the 
PdH\textsubscript{x} compound. Around 40$^\circ$C--50$^\circ$C, the transition region 
begins, and hydrogen desorption is progressively enhanced \cite{sharpe2025simultaneous}. 
Consequently, 43$^\circ$C is the optimal operating temperature, providing maximum hydrogen 
retention and minimal lattice degradation.

W-doped VO\textsubscript{2} is essential for four-state operation, which can be validated 
through Eq.~(10). Calculating the hydrogen desorption rate at 341~K or 68$^\circ$C 
using Eq.~(10) gives $2.41 \times 10^{-5}~\text{s}^{-1}$, which yields a half-life of only 8~hours 
using Eq.~(13). The cumulative hydrogen loss per cycle is then estimated from the 
first-order fractional-conversion expression. Fractional hydrogen loss Eq.~(14) is 
consistent with the exponential decay form $1 - e^{-kt}$ used to describe the fraction 
reacted in first-order transformations in kinetic analyses.
\begin{equation}
    \tau_{1/2} = \frac{\ln 2}{k_\text{des}} \tag{13}
\end{equation}
\begin{equation}
    \text{Fractional H loss} = 1 - \exp(-k_\text{des}\,\Delta t) \tag{14}
\end{equation}
Repeating the same calculation for 43$^\circ$C (316~K), the desorption rate was 5.7 times 
slower than at 68$^\circ$C, extending the half-life to 46~hours and limiting cumulative 
loss to approximately 6.5\% over 1000 cycles \cite{bleu2023towards}.

\subsection{Fabrication Process}

The W-doped VO\textsubscript{2}/Pd MIM plasmonic sensor proposed in this work 
is designed to be realizable through standard top-down nanofabrication 
techniques. The fabrication sequence is depicted in Fig.~\ref{fabrication}, 
and proceeds through five principal stages.

\begin{enumerate}[label=\Roman{enumi}.]

\item \textbf{\textit{Substrate preparation:}} A thermally oxidized silicon 
wafer is used as the starting platform, as shown in 
Fig.~\ref{fabrication}(a). The bilayer stack consists of a bulk Si layer 
for structural rigidity and a thermally grown SiO\textsubscript{2} layer that 
provides optical isolation between the substrate and the active plasmonic 
region above it. Prior to any deposition, the wafer surface is thoroughly 
cleaned to remove organic and particulate contaminants, ensuring strong 
adhesion in subsequent layers \cite{grigorescu2009resists, gangnaik2017new}.

\item \textbf{\textit{Resist coating:}} A hydrogen silsesquioxane (HSQ) 
electron-beam resist is spin-coated uniformly over the wafer surface, as shown 
in Fig.~\ref{fabrication}(b). HSQ is selected as the resist material due 
to its excellent resolution characteristics and its negative-tone behavior 
under electron beam exposure, which makes it well suited for defining the dense 
nanoscale features of the MIM cavity geometry \cite{grigorescu2009resists, 
schmid2004nonaqueous, gangnaik2017new}. A soft-baking step follows to improve 
resist adhesion and uniformity.

\item \textbf{\textit{Electron beam exposure:}} The coated wafer is subjected 
to EBL exposure to define the W-doped VO\textsubscript{2} resonator pattern in 
the resist, as shown in Fig.~\ref{fabrication}(c). EBL is indispensable at 
this stage because the critical dimensions of the sensor including the 
700~nm ring cavity, the 150~nm ring width, and the nanometer-scale coupling 
gaps fall well below the resolution limits of conventional optical 
lithography \cite{grigorescu2009resists, gangnaik2017new, taylor2014electron, 
cinel2012electron}. The focused electron beam writes the pattern with high 
fidelity, enabling precise control over the resonator geometry.

\item \textbf{\textit{Etching and material deposition:}} After resist 
development, the patterned HSQ layer serves as a hard mask for reactive ion 
etching (RIE), which transfers the resonator pattern into the underlying 
W-doped VO\textsubscript{2} film, as shown in 
Fig.~\ref{fabrication}(d). A subsequent EBL step defines the central Pd 
nanobar within the cavity. Pd is then deposited by electron-beam evaporation, 
and a lift-off process removes excess material, leaving behind only the 
precisely positioned nanobar \cite{grigorescu2009resists, gangnaik2017new}. 
It is worth noting that while the Pd nanobar is defined during fabrication, 
the PdH\textsubscript{x} state is only created in situ during actual 
H\textsubscript{2} exposure during operation.

\item \textbf{\textit{Waveguide patterning and final lift-off:}} In the 
concluding fabrication step, the surrounding Pd cladding and the MIM bus 
waveguide are patterned through a final EBL-assisted metal deposition and 
lift-off cycle, as shown in Fig.~\ref{fabrication}(e). This step completes 
the sensor architecture by forming the input and output waveguide arms and 
leaving the ring-shaped sensing cavity, the material under sensing (MUS) 
region, deliberately open \cite{rakib2023oxidation}. This open cavity 
allows the analyte, whether a gas or liquid, to penetrate the sensing volume 
and interact directly with the evanescent plasmonic field confined at the 
metal--dielectric interfaces.

\end{enumerate}

For scalable fabrication, nanoimprint lithography can be considered after making a 
high-resolution master mold. Compared with direct EBL writing, nanoimprint lithography is 
more suitable for repeated large-area fabrication because it is faster and more 
cost-effective. Previous MIM plasmonic sensor studies have suggested EBL, focused ion beam 
lithography, and nanoimprint lithography as possible high-resolution fabrication methods, 
with nanoimprint lithography being more suitable for scalable device production \cite{rakib2023oxidation}.

\section{Result Analysis}

\subsection{Results and Findings}
\label{subsec:results}

\begin{figure*}
	\centering
	\includegraphics[width=.7\textwidth,height=0.82\textheight,keepaspectratio]{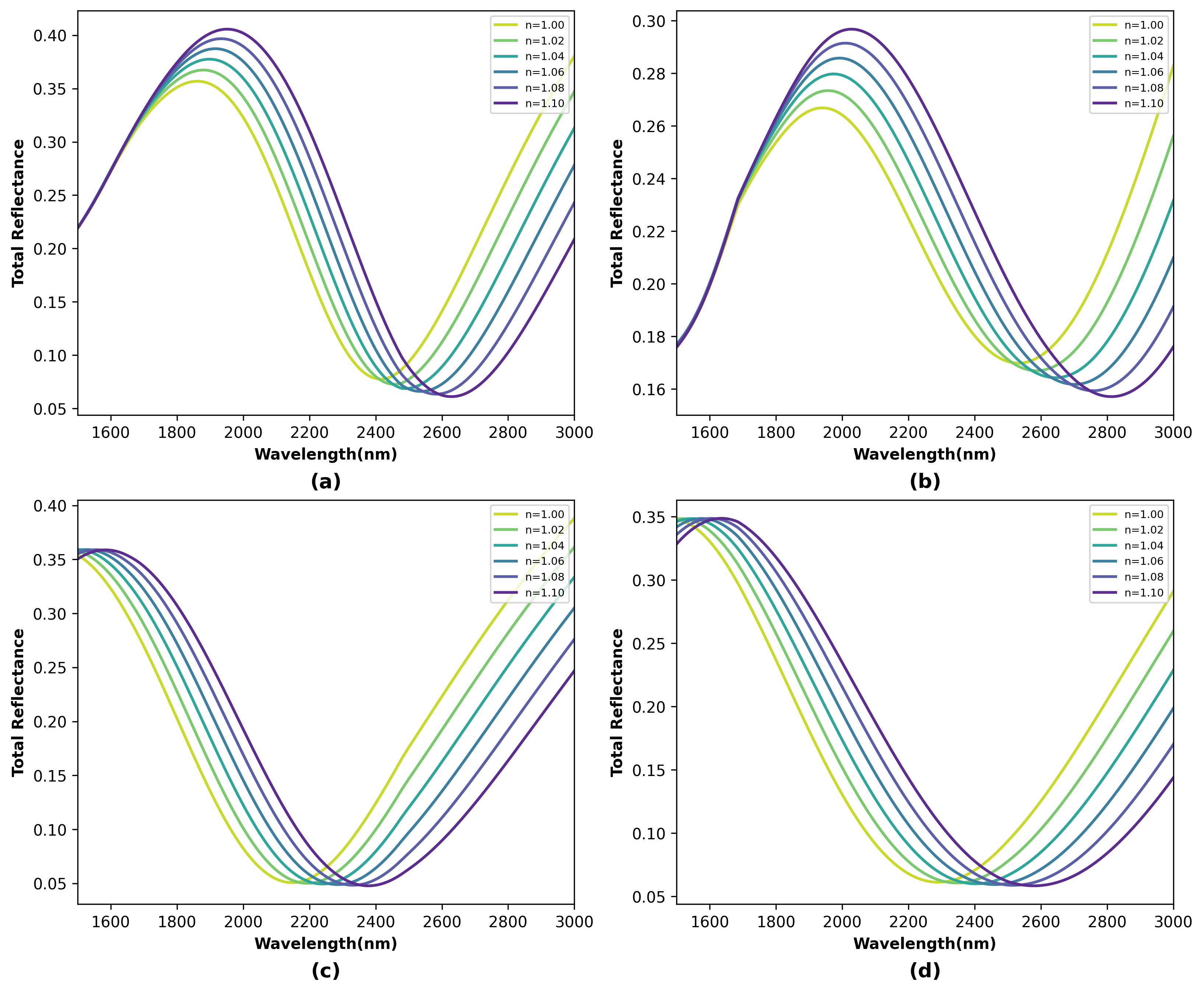}
	\caption{Total reflectance vs.\ wavelength for (a)~State I, (b)~State II, (c)~State III, and (d)~State IV logic for undoped VO\textsubscript{2} and PdH\textsubscript{x}.}
	\label{undopedR}
\end{figure*}

\begin{figure*}
	\centering
	\includegraphics[width=.7\textwidth,height=0.82\textheight,keepaspectratio]{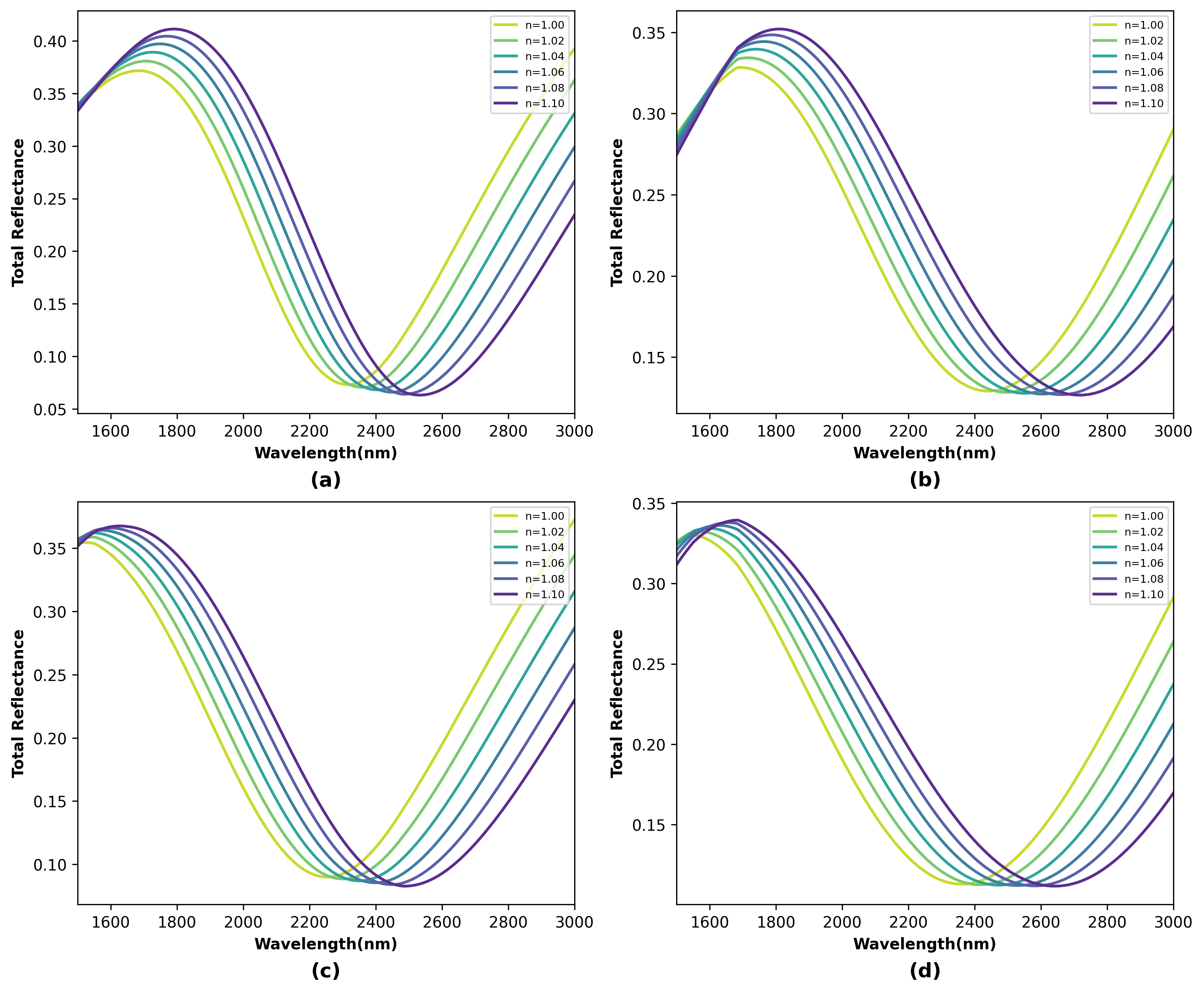}
	\caption{Total reflectance vs.\ wavelength for (a)~State I, (b)~State II, (c)~State III, and (d)~State IV logic for doped VO\textsubscript{2} and PdH\textsubscript{x}.}
	\label{dopedR}
\end{figure*}

The multi-state optical response of the proposed sensor was evaluated for both 
undoped and W-doped VO\textsubscript{2} configurations by sweeping the analyte 
refractive index from 1.00 to 1.10 in increments of 0.02. The reflectance 
spectra for each of the four logic states State~I (00), State~II (01), 
State~III (10), and State~IV (11) are presented in Fig.~\ref{undopedR} 
for the undoped case and Fig.~\ref{dopedR} for the W-doped case. In all 
configurations, a well-defined resonance minimum is observed that shifts 
progressively toward longer wavelengths as the ambient refractive index 
increases, confirming that each state produces a distinct and RI-sensitive 
spectral response.

The collective behavior of all four states is captured in the master curve of 
Fig.~\ref{fig:results_combined}(a), where the reflectance spectra at a fixed RI of 1.00 are 
overlaid for direct comparison. This representation makes explicit the 
bidirectional and mutually opposing nature of the two control stimuli. The 
thermal stimulus, which drives the VO\textsubscript{2} layer from its 
insulating low-$T$ phase toward the lossy high-$T$ phase, produces a 
consistent blue shift in the resonance wavelength. In contrast, the chemical 
stimulus, which drives the $\alpha$-Pd layer toward the $\beta$-PdH\textsubscript{x} 
phase through hydrogen absorption, produces a red shift arising from lattice 
expansion and the associated increase in optical path length. These opposing 
spectral responses establish a robust basis for unambiguous state 
classification: the direction of the resonance shift from the baseline 
immediately identifies whether the active stimulus is thermal or chemical in 
origin.

The base state, State~I ($\alpha$-Pd / Low-$T$ VO\textsubscript{2}), serves 
as the spectral reference. A red-shifted resonance relative to this baseline 
indicates a chemical perturbation, corresponding to a transition to State~II, 
whereas a blue-shifted resonance indicates a thermal perturbation, 
corresponding to a transition to State~III. State~IV also lies to the blue of 
the baseline, since the thermal axis dominates its net shift; however, it is 
identified less by its resonance position than by the elevated reflectance 
level it produces relative to the base state at a comparable wavelength. This 
reflectance-level discrimination complements the wavelength-shift-based 
identification of States~II and~III, providing a three-parameter readout, 
namely shift direction, shift magnitude, and reflectance amplitude, that 
together enable reliable four-state classification.

The six distinct inter-state spectral transitions and their associated physical 
triggers are summarized in Table~\ref{tab:transitioning_states}. The largest transition, 
between State~II and State~III, produces a spectral shift of $-$383~nm, 
reflecting the combined effect of both the chemical and thermal control axes 
acting in opposition. The smallest transition, between State~I and State~II, at $+$116~nm, 
reflects the chemical (hydrogen-absorption) shift acting alone at its 
smallest observed magnitude among the six pairings. These results confirm that the 
four-state sensor architecture maintains 
spectral orthogonality between all states: each state occupies a 
distinguishable region of the reflectance spectrum, which is a prerequisite 
for reliable logic-enabled sensing.

The W-doped case, shown in Fig.~\ref{dopedR}, reproduces the same 
four-state logic structure as the undoped device. Each logic state retains a 
clear, RI-dependent resonance dip, and the directional relationships between 
states are preserved. The preservation of 
the multi-state logic structure under doping confirms that W-doped 
VO\textsubscript{2} is a viable and thermally practical switching material for 
the proposed sensor, without compromising the fundamental spectral 
discrimination capability of the device.
\begin{table}
\centering
\caption{Transitioning States, Physical Triggers, and Resultant Wavelength Shifts}\label{tab:transitioning_states}
\begin{tabular}{lcc}
\toprule
\textbf{Transition} & \textbf{Physical Trigger} & \textbf{Spectral Shift $\Delta\lambda$ (nm)} \\
\midrule
I $\to$ III  & Thermal transition    & $-267$   \\
I $\to$ II   & Chemical transition   & $+116$  \\
I $\to$ IV   & Dual-phase transition & $-126$   \\
III $\to$ IV & Chemical escalation   & $+141$   \\
II $\to$ IV  & Thermal escalation    & $-242$   \\
II $\to$ III & Axis swapping         & $-383$  \\
\bottomrule
\end{tabular}
\end{table}

The four-state configuration generates six distinct spectral transitions governed by two 
independent control axes, thermal and chemical. In Table~\ref{tab:transitioning_states}, 
all spectral shifts and triggers have been shown. This bidirectional response shows that 
the sensor can realize multiple distinct logic states while keeping the spectra for each 
state well separated. In other words, it preserves spectral orthogonality between states, 
which is crucial for reliable next-generation logic-enabled sensing, as explained in the 
application section.

\subsection{Performance Parameters}
Three standard figures of merit are used to 
quantify the spectral sensing performance of the 
proposed four-state platform: the refractive index 
sensitivity~($S$), the figure of merit~(FOM), and 
the resonance quality factor~($Q$)~\cite{shalabney2011sensitivity}. 
Each metric is extracted directly from the simulated 
reflectance spectra by sweeping the ambient refractive 
index across the range of interest.

Among these, $S$ captures the spectral responsivity 
of the resonance dip to incremental changes in the 
surrounding medium. It is defined as:
\begin{equation}
    S = \frac{\Delta\lambda_\text{res}}{\Delta n} \tag{15}
\end{equation}
where $\Delta\lambda_\text{res}$ is the resonance shift 
produced by a refractive index increment $\Delta n$. 
A steeper shift per unit index change reflects 
stronger light--matter interaction in the sensing 
volume.

The FOM further weights sensitivity by resonance 
linewidth, rewarding structures that combine large 
spectral shifts with spectrally narrow dips:
\begin{equation}
    \text{FOM} = \frac{S}{\text{FWHM}} \tag{16}
\end{equation}
where FWHM denotes the full width at half maximum 
of the reflectance minimum. Narrower linewidths 
directly improve the resolvability of closely spaced 
resonant features.

The quality factor $Q$ independently characterizes 
cavity confinement and modal loss, irrespective of 
analyte-induced shifts:
\begin{equation}
    Q = \frac{\lambda_\text{res}}{\text{FWHM}} \tag{17}
\end{equation}
where $\lambda_\text{res}$ is the center wavelength 
of the resonance dip. Elevated $Q$ values are 
indicative of reduced radiative and ohmic damping 
within the plasmonic cavity.

\begin{figure*}
\centering
\begin{subfigure}[b]{0.32\linewidth}
    \centering
    \includegraphics[width=\linewidth,height=0.3\textheight,keepaspectratio]{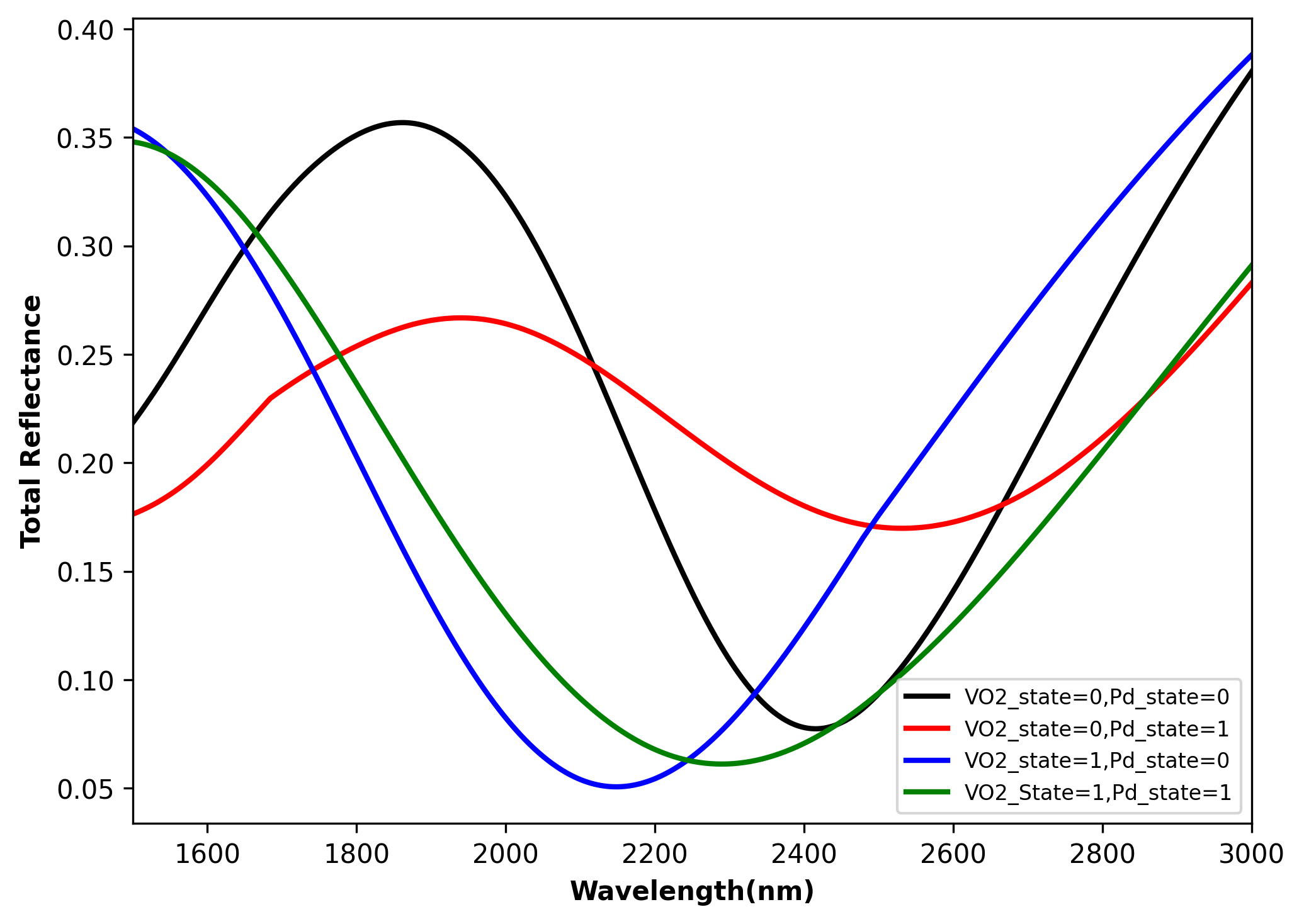}
    \caption{}
    \label{fig:master}
\end{subfigure}
\hfill
\begin{subfigure}[b]{0.32\linewidth}
    \centering
    \includegraphics[width=\linewidth,height=0.3\textheight,keepaspectratio]{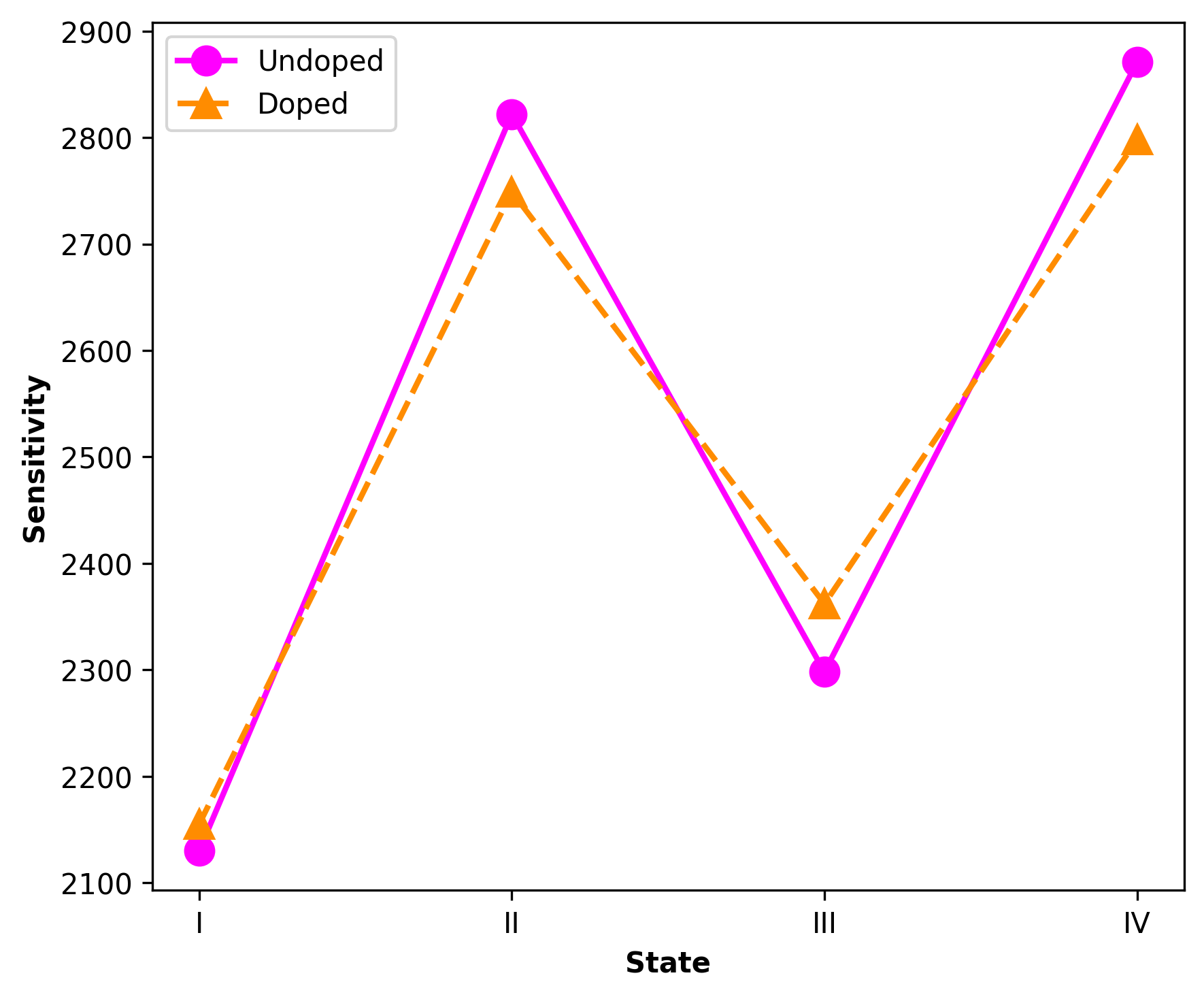}
    \caption{}
    \label{sensitivty}
\end{subfigure}
\hfill
\begin{subfigure}[b]{0.32\linewidth}
    \centering
    \includegraphics[width=\linewidth,height=0.3\textheight,keepaspectratio]{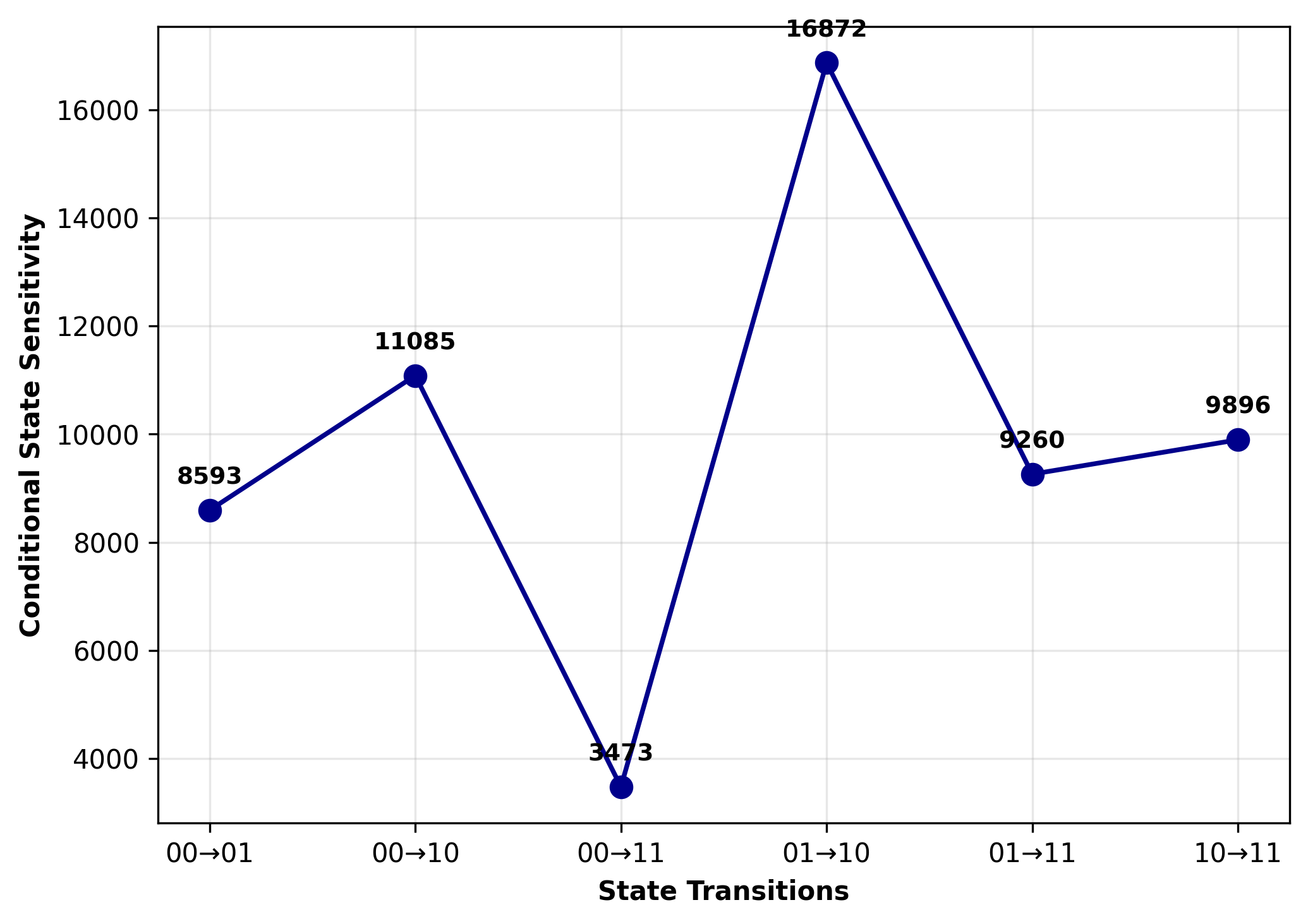}
    \caption{}
    \label{CSS}
\end{subfigure}
\caption{(a) Master curve at RI~=~1.00 for all four states. (b) Sensitivity plot with respect to doped and undoped VO\textsubscript{2} and PdH\textsubscript{x}. (c) Conditional state sensitivity for all six state transitions.}
\label{fig:results_combined}
\end{figure*}

\begin{figure*}
    \centering
    \includegraphics[width=0.85\textwidth,height=0.82\textheight,keepaspectratio]{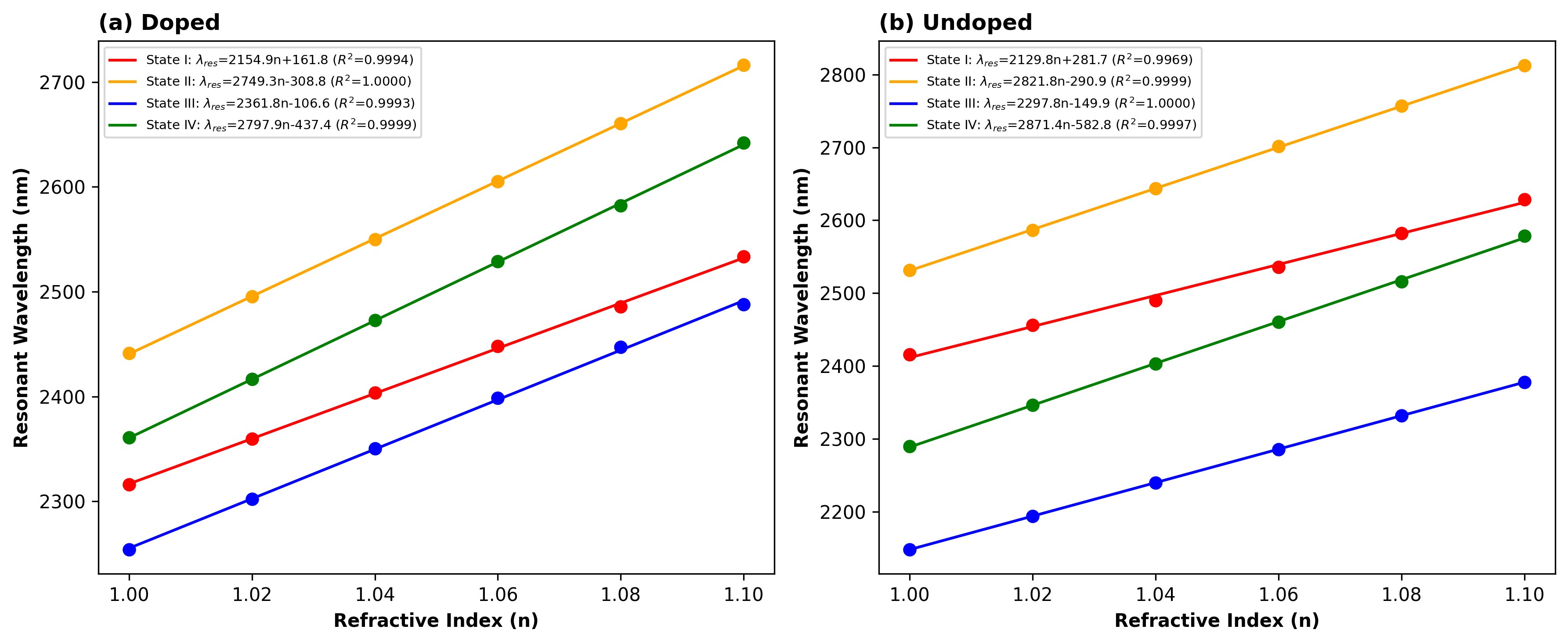}
    \caption{Comparison of linearity curves for both (a) doped and (b) undoped cases across States I--IV.}
    \label{FIG:linear_combined}
\end{figure*}

The spectral sensitivities extracted from Fig.~\ref{undopedR} are 2{,}130, 2{,}822, 
2{,}298, and 2{,}871~nm/RIU for States~I, II, III, and IV of the undoped device, 
respectively. For the doped device, the sensitivities extracted from Fig.~\ref{dopedR} 
are 2{,}155, 2{,}749, 2{,}362, and 2{,}798~nm/RIU for the corresponding states, which are plotted in Fig.~\ref{fig:results_combined}(b). The 
corresponding linear fits of resonant wavelength against analyte refractive index for the 
undoped configuration achieve a coefficient of determination above 0.996 across the full 
1.00--1.10 index range, confirming that the underlying resonance shift is genuinely linear 
over this span, as shown in Fig.~\ref{FIG:linear_combined}. As 
observed in these results, State~II and State~IV exhibit the highest sensitivity among the 
undoped states, with State~IV remaining the single highest of all four. The doped device 
preserves the same relative ordering (State~I lowest, followed by State~III, then State~II, 
with State~IV again the highest), confirming that W-doping does not disturb the underlying 
hierarchy of state discrimination even though it shifts the absolute sensitivity of each 
state individually. This behavior is advantageous for robust multi-state refractive-index 
encoding, as it ensures that the primary state for signal detection remains predictable 
under varying material conditions \cite{shalabney2011sensitivity}.


The calculated results show a maximum figure of merit (FOM) of 
approximately 3.49 for the doped device, whereas the undoped device exhibits a higher value 
of 4.73, with both maxima occurring at State~II. These FOM values are primarily constrained by spectral broadening,
which originates from the intrinsic ohmic losses within the Pd layers of the 
cavity. As shown in Fig.~\ref{pd}, Pd possesses a significantly larger 
imaginary part of the permittivity compared to noble metals like Au or Ag in the 
visible--NIR regime \cite{werner2009optical}, resulting in enhanced damping of surface 
plasmon modes and a consequent reduction in both the quality factor and spectral resolution.


Similar FOM constraints have been documented by Langhammer et al.\ in Pd-based 
metal-hydride sensors, where significant hydrogen-induced wavelength shifts are accompanied 
by broad resonance profiles due to high material damping \cite{wadell2014plasmonic}. 
Furthermore, structural modifications such as alloying or the integration of additional 
layers can introduce extra scattering channels. While alloying may enhance sensitivity, 
the resonance linewidth typically remains broad \cite{wadell2015hysteresis}. Recent 
investigations into Pd-coated nanogratings and ordered Pd nanostructures also report 
high-amplitude wavelength responses coupled with wide resonance bands, highlighting the 
impact of Pd-induced losses on spectral resolution 
\cite{subramanian2020palladium,xu2022large}. For Pd-based plasmonic hydrogen sensors, 
many studies prioritize catalytic and hydrogen-responsive performance (sensitivity, 
detection limit, and response time) over achieving record-high spectral figures of merit, 
accepting a trade-off between resonance sharpness and material functionality 
\cite{sterl2020design,wadell2014plasmonic}.

\subsection{Conditional State Sensitivity}
To comprehensively evaluate the multi-dimensional tuning capabilities of the proposed 
device, we introduce a novel performance metric: \textit{conditional state sensitivity}. 
Unlike traditional plasmonic sensors that exhibit a single, static sensitivity value, this 
reconfigurable sensor demonstrates a dynamic, state-dependent responsivity dictated by the 
interplay of orthogonal thermal and chemical stimuli. This parameter quantifies 
the effective spectral displacement per unit change in refractive index for each specific 
inter-state transition. For an ordered pair of states $A \to B$, CSS is defined using the 
explicit resonance-wavelength functions of the two states as
\begin{equation}
    S_\text{cond}^{A \to B} = 
    \frac{\lambda_\text{res}^{B}(n_0 + \Delta n) - \lambda_\text{res}^{A}(n_0)}{\Delta n} \tag{18}
\end{equation}
where $\lambda_\text{res}^{A}(n_0)$ is the resonant wavelength of state $A$ at the reference 
analyte index $n_0 = 1.00$, $\lambda_\text{res}^{B}(n_0 + \Delta n)$ is the resonant 
wavelength of state $B$ at the perturbed index, and $\Delta n = 0.02$ throughout this work. 
CSS therefore combines the fixed inter-state offset of Table~\ref{tab:transitioning_states} 
with the intra-state refractive-index shift of Eq.~(15), and is not a substitute for the 
conventional single-state sensitivity $S$: it measures how far apart two operating states 
place their resonances under a common index perturbation, which is the quantity that governs 
state-discrimination error rather than index resolution. Figure~\ref{fig:results_combined}(c) reports the 
magnitude $|S_\text{cond}^{A \to B}|$ for all six transitions, consistent with 
Table~\ref{tab:transitioning_states}, which records the signed value; the direction of each 
transition (red-shifting or blue-shifting) is given there.
\begin{figure*}
	\centering
	\includegraphics[width=.75\textwidth,height=0.82\textheight,keepaspectratio]{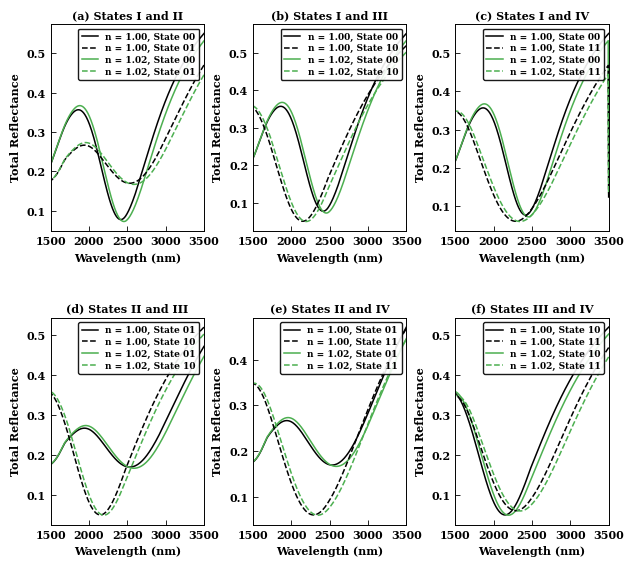}
	\caption{Reflectance plots at RI~=~1.00 and 1.02 for each state-pair transition.}
	\label{conditional_RI}
\end{figure*}

Applying Eq.~(18) to the State~II to State~III transition of 
Fig.~\ref{conditional_RI}(d), the resonance of State~III at the perturbed index 
$n=1.02$ occurs at $\lambda_\text{res}^{III}(1.02) = 2193.8$~nm, while the resonance of 
State~II at the reference index $n=1.00$ occurs at $\lambda_\text{res}^{II}(1.00) = 
2531.2$~nm. The two are separated by 
$|\Delta\lambda_\text{res}| = 337.4$~nm, which over $\Delta n = 0.02$ gives 
$|S_\text{cond}| = 337.4/0.02 = 16{,}872$~nm/RIU. 
The conditional state sensitivity plot (Fig.~\ref{fig:results_combined}(c)) maps 
this multidimensional performance across all six possible switching pathways, and is derived from Fig.~\ref{conditional_RI}. As 
demonstrated in the plot, the sensitivity is highly tunable depending on the active state 
configuration, peaking during the 
transition between State~II (01) and State~III (10), thereby validating the sensor's robust 
multi-state resolving power. Conditional RI sensitivity for other state transitions is also 
significantly high, concluding that each state has very high contrast and can be clearly 
distinguished. Since every inter-state separation in Fig.~\ref{fig:results_combined}(c) exceeds the resonance 
linewidth of the corresponding dips, the four states remain spectrally resolvable at the 
simulated spectral resolution.

\subsection{Performance Benchmarking and Functional Analysis}

\begin{table*}
\caption{Performance comparison of the proposed sensor 
against representative MIM plasmonic refractive index 
sensors from contemporary literature.}
\label{tab:comparison}
\centering
\resizebox{\textwidth}{!}{%
\begin{tabular}{llllll}
\toprule
\textbf{Structure Type} 
& \textbf{Input Stimuli} 
& \textbf{Tunability} 
& \textbf{Logic Capability}
& \textbf{No. of States}
& \textbf{Sensitivity ($S$)} \\
\midrule

Semicircular-Ring MIM~\cite{chau2023enhanced}
& Refractive Index & Static & N/A & 1
& 2{,}900~nm/RIU \\[4pt]

Asymmetric Fano-MIM~\cite{wang2019fano}
& Refractive Index & Static & N/A & 1
& 1{,}114.3~nm/RIU \\[4pt]

Dual Nanorod Array MIM~\cite{rakhshani2021refractive}
& Refractive Index & Static & N/A & 1
& 1{,}090~nm/RIU \\[4pt]

Dual-Stub Fano-MIM~\cite{qu2024mim}
& Refractive Index & Static & N/A & 1
& 3{,}580~nm/RIU \\[4pt]

SRRR Baffled MIM~\cite{chen2024srrr}
& Refractive Index & Static & N/A & 1
& 2{,}560~nm/RIU \\[4pt]

Triple-Disk T-Cavity MIM~\cite{yang2024nano}
& Refractive Index & Static & N/A & 1
& 3{,}020~nm/RIU \\[4pt]

Multichannel Elliptical-Ring MIM~\cite{chouChao2023}
& Refractive Index & Static & N/A & 1
& 4{,}500~nm/RIU \\[4pt]

\textbf{Orthogonal MIM 
(W-VO\textsubscript{2}~+~Pd) [This work]}
& \textbf{Dual (Thermal~+~Chemical)} 
& \textbf{Active} 
& \textbf{4-State Logic} 
& \textbf{4}
& \textbf{2{,}798~nm/RIU (W-doped); 2{,}871~nm/RIU (undoped reference)} \\

\bottomrule
\end{tabular}%
}
\end{table*}
Table~\ref{tab:comparison} systematically benchmarks the 
proposed orthogonal W-VO\textsubscript{2}/Pd MIM architecture 
against seven representative plasmonic refractive-index sensors 
from contemporary literature. Among the static single-state 
designs, Chau et al.~\cite{chau2023enhanced} demonstrated a 
semicircular-ring MIM resonator with a sensitivity of 
2{,}900~nm/RIU, while Wang et al.~\cite{wang2019fano} and 
Rakhshani et al.~\cite{rakhshani2021refractive} reported 
asymmetric Fano-MIM and dual nanorod array configurations 
achieving 1{,}114.3~nm/RIU and 1{,}090~nm/RIU, respectively. 
More recent works have pushed static single-state sensitivity 
further: Qu et al.~\cite{qu2024mim} demonstrated an 
independently tunable dual-stub Fano-MIM structure reaching 
3{,}580~nm/RIU, Yang et al.~\cite{yang2024nano} reported a 
triple-disk T-shaped cavity sensor at 3{,}020~nm/RIU, Chen 
et al.~\cite{chen2024srrr} achieved 2{,}560~nm/RIU with a 
semicircular ring rectangular resonator, and Chou Chao 
et al.~\cite{chouChao2023} attained the highest static 
sensitivity in this comparison at 4{,}500~nm/RIU using a 
multichannel elliptical-ring geometry. Despite their 
individually strong sensitivities, all seven benchmarked 
structures share a fundamental constraint: each operates as 
a single-state device driven exclusively by refractive index 
variation, with no provision for active spectral 
reconfiguration or multi-state logic encoding.

The present work extends this landscape along both the 
performance and functionality axes simultaneously. By 
embedding W-doped VO\textsubscript{2} and Pd within an 
orthogonal MIM cavity, the proposed device supports two 
fully decoupled control channels --- chemical stimulus 
governing the Pd phase and thermal stimulus governing the 
VO\textsubscript{2} phase --- that together access a 
deterministic four-state logic space unavailable to any 
single-stimulus architecture. Despite integrating these 
active materials, the practical W-doped device achieves a peak 
refractive-index sensitivity of 2{,}798~nm/RIU in the State~IV 
configuration, rising to 2{,}871~nm/RIU for the undoped 
reference configuration, and therefore remains 
competitive with the majority of the benchmarked 
structures and within a factor of two of the highest reported 
static value. More critically, the proposed sensor introduces 
conditional state sensitivity~(CSS) as a novel figure of 
merit, reaching a peak value of 16{,}872~nm/RIU across the 
State~II-to-State~III transition --- a metric with no 
meaningful equivalent in single-state sensors and therefore 
not directly comparable to any benchmarked entry. The 
convergence of competitive RI sensitivity, orthogonal 
dual-stimuli addressability, and four-state logic capability 
establishes the proposed platform as a qualitatively distinct 
advancement over all seven benchmarked designs: whereas 
those works optimize a single spectral response, the proposed 
device encodes four unambiguously distinguishable states, a 
capability that static single-stimulus sensors are 
architecturally unable to replicate.

\section{Applications}

\subsection{Multistate Environmental and Safety Monitoring for Industrial Autonomy}

\begin{figure}
	\centering
	\includegraphics[width=0.9\linewidth,height=0.44\textheight,keepaspectratio]{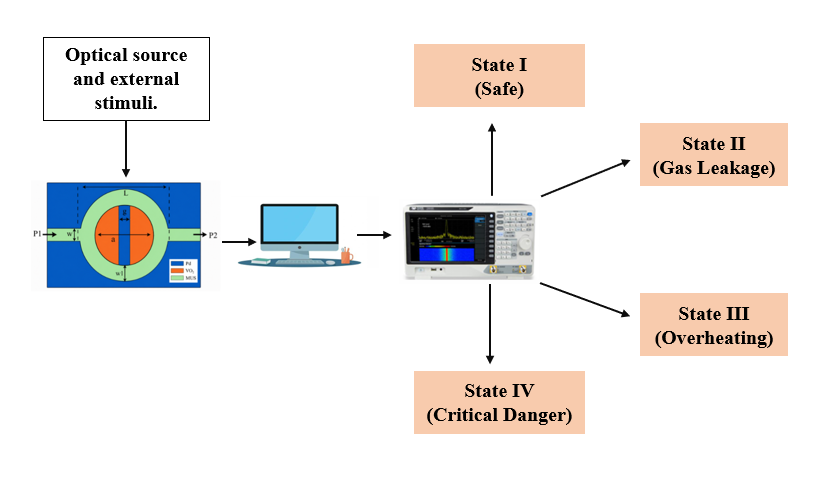}
	\caption{State Decision Flow Chart for multistate safety monitoring.}
	\label{app1}
\end{figure}

The four distinct resonance states can be reinterpreted as a comprehensive framework for 
multistate environmental and safety monitoring, as illustrated in the flow diagram of 
Fig.~\ref{app1}. While traditional sensors often provide a simple binary alarm, the 
proposed W-doped VO\textsubscript{2}/Pd platform generates a nuanced spectral code that 
classifies the severity of environmental hazards in real-time. The system's output is 
governed by the simultaneous monitoring of thermal and chemical thresholds, providing a 
``hardware-level'' classification of safety conditions.

By mapping the permutations of temperature and hydrogen presence, the device establishes 
a hierarchical safety protocol based on unique spectral signatures:

\begin{itemize}
    \item \textbf{Safe State (Low-T $+$ $\alpha$-Pd):} Under standard operating conditions 
    (e.g., 25$^\circ$C and no H\textsubscript{2}), VO\textsubscript{2} remains in its 
    Low-$T$ optical phase, and the Pd layer remains stable. This produces a baseline resonance 
    signaling State~I, a secure environment.

    \item \textbf{Chemical Warning (Low-T $+$ $\beta$-PdH\textsubscript{x}):} In the 
    event of a hydrogen leak at ambient temperatures, the formation of palladium hydride 
    (PdH\textsubscript{x}) significantly alters the complex refractive index of the sensing 
    layer. This triggers a distinct spectral shift to State~II, providing a dedicated 
    warning for chemical contamination without thermal complications.

    \item \textbf{Thermal Warning (High-T $+$ $\alpha$-Pd):} If the environment reaches 
    the critical threshold of 43$^\circ$C, the W-doping facilitates a transition of the 
    VO\textsubscript{2} layer from its Low-$T$ optical phase to its High-$T$ optical 
    phase. This transition dramatically modifies the plasmonic 
    resonance condition, shifting the dip to State~III and signaling an overheating hazard.

    \item \textbf{Critical Danger State (High-T $+$ $\beta$-PdH\textsubscript{x}):} The 
    most severe condition occurs when both overheating and hydrogen leakage happen 
    concurrently. This dual-stimulus state produces the unique resonance signaling 
    State~IV, which serves as a high-priority ``critical danger'' trigger for emergency 
    shutdown procedures.
\end{itemize}

As discussed previously, states can be identified from red shift or blue shift from the 
base state for chemical warning and thermal warning, respectively, and reflectance level 
increases from the baseline state for critical danger states; therefore, real-time 
monitoring is possible.


\subsection{Optical Logic Gate for Intelligent Decision-Making}

\begin{figure}
	\centering
	\includegraphics[width=0.9\linewidth,height=0.44\textheight,keepaspectratio]{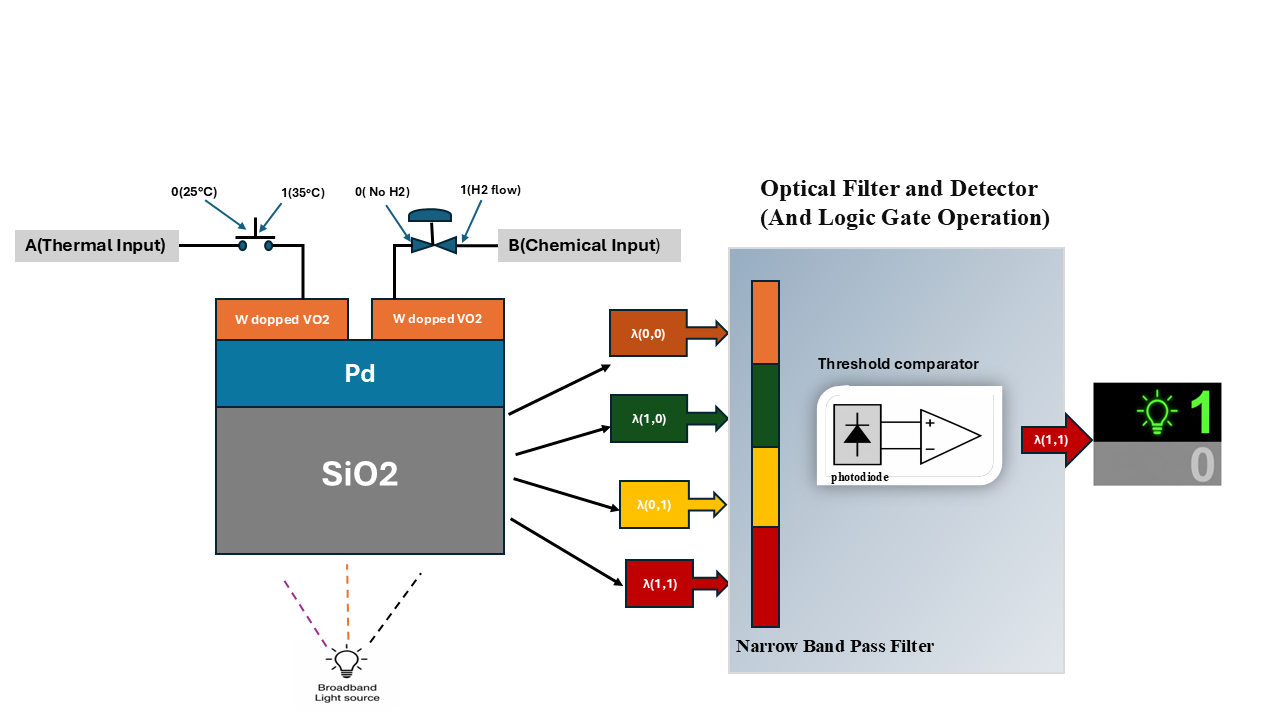}
	\caption{ Four-state sensor as optical AND logic gate.}
	\label{app2}
\end{figure}

The primary application of the proposed W-doped VO\textsubscript{2}/Pd resonator is as a 
functional optical logic gate with integrated sensing. In the configuration illustrated in 
Fig.~\ref{app2}, the device implements an AND operation with two independent physical 
inputs. Input~A (thermal) is encoded as a binary signal, where 0 denotes the 
low-temperature state (e.g., 25$^\circ$C, VO\textsubscript{2} in its Low-$T$ optical phase) 
and 1 denotes the high-temperature state (e.g., 43$^\circ$C, VO\textsubscript{2} in its 
High-$T$ optical phase). Input~B 
(chemical) is defined by the hydrogen environment: 0 for no H\textsubscript{2} (Pd) and 1 
for H\textsubscript{2} flow leading to PdH\textsubscript{x} formation. Under broadband 
illumination, the four input combinations $(A, B) = (0,0),\ (0,1),\ (1,0),$ and $(1,1)$ 
generate four distinct resonance wavelengths at each state.

To realize an optical AND gate, a narrow band-pass filter is centered at State~IV, which 
appears only when both inputs are ``1'' (high temperature and hydrogen present). A 
photodiode is employed as a minimal transducer to convert this wavelength-selective optical 
output into an electrical signal, enabling direct interfacing with conventional read-out 
and control electronics. The resulting photocurrent is compared against a fixed reference 
by a simple threshold comparator, which digitizes the analog signal: when the photocurrent 
exceeds the threshold, the comparator output switches to logic ``1''; otherwise, the output 
remains ``0.'' This scheme provides a robust binary decision that is insensitive to small 
intensity fluctuations and faithfully reflects the underlying four-state logic.

This purely optical implementation of an AND operation demonstrates that the four-state 
VO\textsubscript{2}/Pd sensor can function as an integrated logic element, in which 
external thermal and chemical stimuli are directly converted into digital information 
through discrete resonance wavelengths. By eliminating the need for intermediate electronic 
encoding or complex signal processing, the device realizes true sensing in the optical 
domain, where the environment itself programs the logical state of the system. Such a 
capability is highly attractive for future on-chip optical computing, smart photonic 
circuitry, and distributed hardware-level decision making in safety-critical platforms, 
autonomous monitoring networks, and other scenarios where compact, low-power, and 
self-deciding photonic components are required.

\begin{table*}
    \centering
    \caption{Logic truth table for the four-state plasmonic system.}
    \label{tab:truth_table}
    \begin{tabularx}{\textwidth}{>{\centering\arraybackslash}X 
    >{\centering\arraybackslash}X >{\centering\arraybackslash}X 
    >{\centering\arraybackslash}X >{\centering\arraybackslash}X}
        \toprule
        \textbf{Input A (Temp)} & \textbf{Input B (H$_2$)} & 
        \textbf{Resonance $\lambda$} & \textbf{Filter Output} & 
        \textbf{Logic Out} \\
        \midrule
        0 (Low)  & 0 (None)       & $\lambda_\text{I}$   & Low  & 0 \\[10pt]
        0 (Low)  & 1 (H$_2$ flow) & $\lambda_\text{II}$  & Low  & 0 \\[10pt]
        1 (High) & 0 (None)       & $\lambda_\text{III}$ & Low  & 0 \\[10pt]
        1 (High) & 1 (H$_2$ flow) & $\lambda_\text{IV}$  & High & 1 \\
        \bottomrule
    \end{tabularx}
\end{table*}

The performance of the proposed four-state plasmonic system as an optical AND logic gate 
is evaluated using two key figures of merit: contrast ratio (CR) and modulation depth 
(MD). These parameters are widely adopted in the characterization of plasmonic logic gates 
and optical switching devices to quantify the discrimination capability between ON and OFF 
states \cite{haque2024optical,abdulwahid2023multiple}.

The contrast ratio is defined as the logarithmic ratio of the output power in the ON state 
to that in the OFF state, measured at the AND gate operating wavelength:
\begin{equation}
    \text{CR} = 10\log_{10}\!\left(\frac{P_\text{ON}}{P_\text{OFF}}\right) \tag{19}
\end{equation}
where $P_\text{ON}$ and $P_\text{OFF}$ denote the optical power at the output for the ON 
(State~IV) and OFF (States~I, II, and III) conditions, respectively. A higher CR value 
indicates more robust binary discrimination between logical states \cite{haque2024optical}.

The modulation depth quantifies the relative change in output intensity between the ON and 
OFF states \cite{haque2024optical,abdulwahid2023multiple}:
\begin{equation}
    \text{MD} = \frac{R_\text{ON,max} - R_\text{OFF,min}}{R_\text{ON,max}} 
    \times 100\% \tag{20}
\end{equation}
where $R_\text{ON,max}$ is the maximum reflectance in the ON state and $R_\text{OFF,min}$ 
is the minimum reflectance among all OFF states at the operating wavelength. At the AND 
gate operating wavelength of 2{,}360.6~nm, corresponding to the doped State~IV resonance at 
$n=1.00$, the proposed system achieves a contrast ratio of 1.66~dB and a modulation depth 
of 31.8\%, indicating functional but modest AND gate discrimination. 
The relatively moderate contrast between State~III and State~IV reflects the comparable 
near-infrared optical response of Pd and PdH\textsubscript{x}, which is an inherent 
characteristic of hydrogen-sensitive plasmonic materials operating in this spectral regime 
\cite{yarahmadi2015subwavelength}.


\section{Conclusion}
\label{sec:conclusion}

In this work, we introduced and numerically investigated a reconfigurable four-state 
plasmonic refractive index sensor based on orthogonal thermal and chemical stimuli, where 
palladium (Pd) and tungsten-doped vanadium dioxide (W-doped VO\textsubscript{2}) were 
integrated into a metal-insulator-metal (MIM) waveguide-cavity structure. The optical 
response of the structure was analyzed using the finite element method (FEM), where 
TM-polarized excitation, port boundary conditions, scattering boundaries, and a refined 
triangular mesh were employed to accurately capture the surface plasmon polariton behavior 
and resonance characteristics. To realize the four-state operation, the Pd/PdH transition 
and the low-T/high-T states of W-doped VO\textsubscript{2} were combined in a controlled 
manner, allowing the sensor to operate in four distinct regimes. The proposed sensor 
achieved a maximum conditional refractive index sensitivity of 16{,}872~nm/RIU and a peak 
sensitivity of 2{,}798~nm/RIU for the W-doped device (2{,}871~nm/RIU for the undoped 
reference configuration), demonstrating effective multi-state discrimination and 
strong sensing performance. Overall, the findings of the study establish the compact and 
numerically evaluated MIM plasmonic sensor for orthogonally controlled sensing, with 
potential applications in multi-analyte detection, industrial monitoring, and logic-enabled 
photonic devices.

\section*{Declaration of Competing Interest}
The authors declare that they have no known competing financial interests 
or personal relationships that could have appeared to influence the work 
reported in this paper.

\section*{Data Availability}
No data was used for the research described in this article.

\section*{Funding}
This study was conducted without the benefit of any external financial 
support.
\section*{Declaration of Generative AI and AI-assisted technologies in the writing process}
During the preparation of this work, the authors used AI tools only to improve the grammar and language of the manuscript. After using this tool, the authors reviewed and edited the content as needed and take full responsibility for the content of the publication.

\bibliographystyle{model1-num-names}
\bibliography{reference}
\end{document}